\documentclass[a4paper]{article}

 \pdfoutput=1

\usepackage{authblk}
\usepackage{charter}
\usepackage{helvet}
\usepackage[T1]{fontenc}
\usepackage[latin9]{inputenc}
\usepackage[a4paper]{geometry}
\usepackage{xcolor}
\usepackage{mathrsfs}
\usepackage{mathtools}
\usepackage{multirow}
\usepackage{amsmath}
\usepackage{amssymb}
\usepackage{graphicx}
\usepackage{setspace}
\usepackage[unicode=true,
 bookmarks=false,
 breaklinks=false,pdfborder={0 0 1},backref=false,colorlinks=false]
 {hyperref}
\usepackage{enumitem}
\makeatletter

\providecommand{\tabularnewline}{\\}

\usepackage{xcolor}
\usepackage{lineno}
\modulolinenumbers[1]
\usepackage{lscape}

\usepackage{savesym}
\numberwithin{equation}{section}
\usepackage{caption}
\usepackage{subcaption}
\usepackage{verbatim}
\usepackage{epsfig}
\usepackage{asymptote}
\graphicspath{{Figures/}} 

\savesymbol{amalg}
\usepackage{mathabx}
\restoresymbol{MAT}{amalg}

\savesymbol{ulcorner}
\savesymbol{urcorner}
\savesymbol{llcorner}
\savesymbol{lrcorner}
\usepackage{amsfonts}
\restoresymbol{AMS}{ulcorner}
\restoresymbol{AMS}{urcorner}
\restoresymbol{AMS}{llcorner}
\restoresymbol{AMS}{lrcorner}
\usepackage[bbgreekl]{mathbbol}
\DeclareSymbolFontAlphabet{\mathbbm}{bbold}
\DeclareSymbolFontAlphabet{\mathbb}{AMSb}

\usepackage{amsthm}
\usepackage{cases}
\usepackage{algorithm}
\usepackage{algpseudocode}
\usepackage{soul}
\usepackage{pict2e}
\usepackage{mwe}
\usepackage{footmisc}
\usepackage[cal=boondox,scr=boondoxo]{mathalfa}
\usepackage{multirow}
\usepackage{array}
\usepackage{colortbl}

\usepackage{todonotes}
\usepackage{lineno}

\usepackage{enumitem}

\usepackage{url}
\usepackage{mwe}

    \AfterLastShipout{\immediate\write\@mainaux{\relax}}%

\newif\ifHNNdraft
\HNNdrafttrue

\newcommand\footnoteref[1]{\protected@xdef\@thefnmark{\ref{#1}}\@footnotemark}

\usepackage{nomencl}
\makenomenclature

\renewcommand{\u}[1]{\boldsymbol{#1}}

\newcommand{\m}[1]{\mathbb{#1}}
\renewcommand{\c}[1]{\mathcal{#1}}
\newcommand{\lsc}[2][\mathscr{l}]{{}^{ #1 }\! #2}

\newcommand{\dsf}[1]{\Delta\boldsymbol{\sf #1}}
\newcommand{\tpsb}[1]{\left. #1 \right.^{\sf T}}
\newcommand{\tipsb}[1]{\left. #1 \right.^{-\sf T}}
\newcommand{\tps}[1]{\left( #1 \right)^{\sf T}}

\newcommand{\hpsb}[1]{\left. #1 \right.^{\sf H}}
\newcommand{\usf}[1]{\u{\sf #1}}
\newcommand{\busf}[1]{\overline{\usf{ #1}}}

\newcommand{\pr}[1]{\left( #1 \right)}
\newcommand{\prb}[1]{\left. #1 \right.}

\newcommand{\newtag}[2]{#1\def\@currentlabel{#1}\label{#2}}
\newcommand{\cu}[1]{\check{\u{ #1}}}
\newcommand{\cusf}[1]{\cu{\sf #1}}
\newcommand{\set}[2]{\left\{#1\, \big|\, #2\right\}} 
\newcommand{\nsd}{{\rm n}_{\rm sd}}
\newcommand{\norm}[1]{\lVert #1\rVert}

\definecolor{carmine}{rgb}{0.59, 0.0, 0.09}

\newtheorem{lemma}{\bf Lemma}[section]

\makeatother
\begin{document}

\title{Determining rigid body motion from accelerometer data through
the square-root of a negative semi-definite tensor, with applications
in mild traumatic brain injury}
\author[1]{Yang Wan}
\author[1]{Alice Lux Fawzi}
\author[1,*]{Haneesh Kesari}
\affil{Brown University School of Engineering, 184 Hope St., Providence, RI, USA}
\affil[*]{Corresponding author, haneesh\_kesari@brown.edu}
\maketitle

\bibliographystyle{elsarticle-num}


\begin{abstract}
Mild Traumatic Brain Injuries (mTBI) are caused by violent head motions
or impacts. Most mTBI prevention strategies explicitly or implicitly
rely on a ``brain injury criterion''.
A brain injury criterion takes some descriptor of the head's motion
as input and yields a prediction for that motion's potential for causing
mTBI as the output. The inputs are descriptors of the head's motion
that are usually synthesized from accelerometer and gyroscope data.
In the context of brain injury criterion the head is modeled as a
rigid body. We present an algorithm for determining the complete motion
of the head using data from only four head mounted tri-axial accelerometers.
In contrast to inertial measurement unit based algorithms for determining
rigid body motion the presented algorithm does not depend on data
from gyroscopes; which consume much more power than accelerometers.
Several algorithms that also make use of data from only accelerometers
already exist. However, those algorithms, except for the recently
presented AO-algorithm [Rahaman MM, Fang W, Fawzi AL, Wan Y, Kesari
H (2020):\textit{ J Mech Phys Solids} 104014], give the rigid body's
acceleration field in terms of the body frame, which in general is
unknown. Compared to the AO-algorithm the presented algorithm is much
more insensitive to bias type errors, such as those that arise from
inaccurate measurement of sensor positions and orientations.
\end{abstract}


\section{Introduction}
\label{sec:intro}
Mild Traumatic Brian Injury (mTBI) is the most common injury among
military personnel and it is estimated that as many as 600 per 100,000
people experience mTBIs each year across the world \cite{cdc2013report,cassidy2004}.
Mild Traumatic Brain Injuries are caused by violent head motions, that may occur from intense blunt impacts to the head in contact sports, motor vehicle accidents, falls following blasts, etc.
In mTBI, the motion of the head causes the soft tissue of the brain
to deform. The magnitude
and time rate of brain deformation can cause brain cells to die \cite{ganpule2017three,bain2000nonrbd,meaney2014mechanics,kleiven2013most,bar2016strain}.

There have been many strategies aimed at preventing mTBI.
In sports, new rules aim to modify player behavior in order to
decrease or eliminate exposure to blunt impacts \cite{kriz2019effect}.
Helmets and neck collars are examples of equipment that can alter
the motion experienced by the head.
A jugular vein compression collar aims to change the stiffness of the
brain making it less susceptible to injury \cite{mannix2020internal}.

Most mTBI prevention strategies explicitly or implicitly rely on a ``brain injury criterion'' for their effective
synthesis, implementation, and evaluation.
A brain injury criterion takes some descriptor of the head's motion
as input and yields a prediction for that motion's potential for causing
mTBI as the output.

When we refer to any aspect of the head's motion, we, in fact, are
referring to that aspect as it pertains to the skull's motion; since
it is the skull's motion that is, at least currently, observable
and quantifiable in the field, either using video recording equipment
or inertial sensor systems.
The Young's modulus of bone from human skulls generally lies in the
2--13 GPa range \cite{Mcelhaney1970,Delille2007,Motherway2009}. In
comparision, brain tissue is extremely compliant.
Recent indentation tests on brain slices that were kept hydrated show
that the Young's modulus of brain tissue lies in the 1--2 kPa range
(white matter $1.9\pm 0.6$ kPa, and gray matter $1.4\pm 0.3$ kPa)
\cite{Budday2015}.
Due to this large disparity between the skull's and the brain's stiffnesses,
in biomechanical investigations of mTBI the skull is usually modeled
as a rigid body \cite{Willinger1999,Wright2013}.
Thus, inputs to the brain injury criteria are rigid body motion descriptors,
such as angular velocity time series, translational acceleration times
series, etc., or a combination of such time series.

Rigid body motion can be thought of as a composition of translatory
and rotatory motions. In initial brain injury criteria
the focus was on the head's translatory motion. Two
of the first published injury criteria are the Gadd Severity Index
(SI) and the Head Injury Criterion (HIC) \cite{gadd1966use,chou1974analytical}.
Both SI and HIC ignore the head's rotations and take
the head's translational acceleration as their input. Later,
however, it was realized that in the context of mTBI the head's rotations
play an even more important role in causing injury than its translations. The
first brain injury criterion to take the rotational aspect of the
head's motion into consideration was GAMBIT \cite{newman1986generalized}. The
input to GAMBIT is the tuple of center-of-mass-acceleration and angular-acceleration
time series. Following the development of GAMBIT, brain
injury criteria that use descriptors that only depend on the head's
rotational motion as inputs have also been put forward. One
such criteria is the Brain Injury Criteria (BrIC) \cite{takhounts2013development}.
Aiming to compliment HIC, BrIC only uses the head's angular velocity
time series as input. We also note that there is currently
significant activity in applying finite element modeling using 2D/3D
anatomically consistent discrete geometry head models to evaluate
or develop new brain injury criteria \cite{Laksari2020,Gabler2018,Gabler2019}.

Irrespective of which existing, or yet to be developed,
brain injury criterion will be used in the future, its successful application will hinge on the availability of a robust algorithm for constructing the motion descriptor that the criterion takes as input from easily measurable data.
Currently, different algorithms are used to obtain the descriptors taken by the injury criteria as inputs.
The inputs to GAMBIT can be obtained from the measurements of one tri-axial
accelerometer and one tri-axial gyroscope mounted in a mouthguard \cite{Hernandez2015}
if the center-of-mass-acceleration and angular-acceleration are obtained
by processing the data using the algorithm in \cite{Camarillo2013}.
In another example the input to BrIC (i.e., angular velocity) is
prepared by numerically integrating the angular acceleration, which
is determined by applying the 6DOF algorithm \cite{Chu2006} to
the data from 12 single-axis accelerometers mounted in a helmet \cite{Rowson2009}. Interestingly
the inputs to most of the currently employed brain injury criteria
can be prepared from the knowledge of a few key rigid body motion
descriptors. To make this idea more concrete, consider
the following equation, which is often used to describe rigid body
motion,

\begin{equation}
\usf{x}(\tau)=\usf{Q}(\tau)\usf{X}+\usf{c}(\tau).
\label{eq:NondimensionalRotationDeformationMapping}
\end{equation}

In $\eqref{eq:NondimensionalRotationDeformationMapping}$ $\tau$ is
a real number that denotes a non-dimensional time instant; $\usf{X}$
is a column matrix of real numbers that denotes the initial position
vector of a rigid body material particle; $\usf{x}(\tau)$ is the
column matrix of real numbers that denotes that material particle's
position vector at the time instance $\tau$; $\usf{Q}(\tau)$ is
a time dependent square matrix of real numbers with positive determinant
whose transpose equals its inverse; and $\usf{c}(\tau)$ is a time
dependent column matrix of real numbers. The matrix
$\usf{Q}(\tau)$ quantifies the rotation or orientation of the rigid
body at the time instance $\tau$, while $\usf{c}(\tau)$ quantifies
the rigid body's translation at that time instance.
The inputs to most current brain injury criteria can be computed from
a knowledge of the maps $\usf{Q}$ and $\usf{c}$ and their first
and second-order time derivatives, i.e., $\usf{Q}'$, $\usf{c}'$, $\usf{Q}''$, $\usf{c}''$.
In this manuscript we present an algorithm for determining these
maps and their derivatives using data from only four tri\textcolor{black}{-axial
accelerometers.}

The presented algorithm has some similarities to the one recently
presented by Rahaman \textit{et al.} \cite{rahaman2019}\footnote{A graphical user interface for applying the AO algorithm to different types of data sets and visualizing its results is freely available \cite{AOAPP2021}}, which is referred to as the AO (accelerometer-only) algorithm. For reasons that will
become clear shortly, we refer to the algorithm that we present in
this manuscript as the $\sqrt{\text{AO}}$-algorithm. The AO-algorithm also presents a framework for completely determining the rigid body's motion, i.e., for constructing the maps $\usf{Q}$ and $\usf{c}$
and their time derivatives, using data only from four tri-axial accelerometers.
The $\sqrt{\textrm{AO}}$-algorithm has all the advantages
of the AO-algorithm.

There are existing algorithms for completely determining a rigid
body's motion using sensor data. However, these algorithms
take data from sensor systems called inertial measurement units (IMUs).
These units contain one or more gyroscopes. One of
the primary advantages of the AO and $\sqrt{\text{AO}}$ algorithms
is their non-dependence on gyroscopes. For a detailed
discussion on why accelerometers are preferable over gyroscopes in
the context of mTBI please see $\S$1 in \cite{rahaman2019}.
Briefly, gyroscopes' power requirements are much higher than those
of accelerometers (gyroscopes consume approximately 25 times more
power than accelerometers \cite{Lee2019}), and algorithms that
aim to construct descriptors of a rigid body's acceleration
using data from gyroscopes add a significant amount of noise
to those descriptors \cite{ovaska1998noise,alonso2005noise}.

Several algorithms exist for constructing inputs to brain injury criteria
that that too only make use of data from accelerometers (Padgaonkar~\textit{et
al.}~\cite{padgaonkar1975measurement}, Genin~\textit{et al.}~\cite{genin1997} and Naunheim~\textit{et al.}~\cite{naunheim2003linear}).
These algorithms, however, give much more limited information than
is given by the AO and the $\sqrt{\text{AO}}$ algorithms. For
example, all these algorithms give the rigid body's acceleration field
in terms of the body frame, which is a set of vectors that are attached
to the rigid body, and hence move with it. These algorithms
do not provide any information of how the body frame is oriented in
space. However, that information is critical for constructing
inputs for the upcoming finite element based brain injury criteria.
The AO and the $\sqrt{\text{AO}}$ algorithms provide complete information
of how the body frame is oriented in space. See $\S$1
in \cite{rahaman2019} for further discussion on the advantages of
the AO and the $\sqrt{\text{AO}}$ algorithms over other algorithms
that also make use of only accelerometer data.

\begin{figure}[ht]
\centering{}
\includegraphics[width=0.85\textwidth]{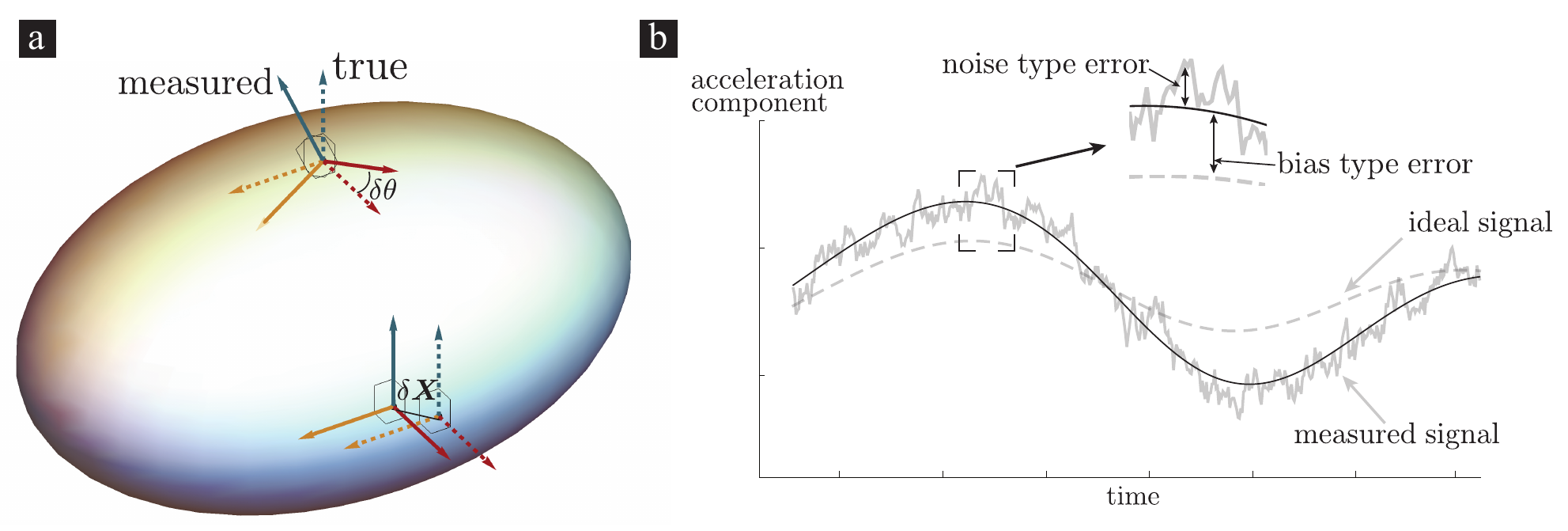}
\caption{Bias and noise type errors in acceleration component measurements. When there are errors $\delta\u{X}$ and $\delta\theta$ in accurately defining an accelerometer's position and orientation, respectively (see (a)), bias type errors can occur in the measurement of acceleration components (see (b)). Noise type errors in the acceleration component measurements are usually a consequence of seismic, electrical, and other types of noise.}
\label{fig:errors}
\end{figure}

Despite its many advantages we note that the AO-algorithm has one
critical limitation. It is quite sensitive to bias type errors in
the accelerometer data. Bias type errors are distinct from random
errors in that they do not arise as a consequence of stochastic processes.
For accelerometers, bias type errors can arise as a consequence of inaccurately defining sensor position and orientation (see Fig.~\ref{fig:errors}).
As we explain below, the advantage of the $\sqrt{\text{AO}}$-algorithm
over the AO-algorithm is that it is far less sensitive to bias type
errors than the AO-algorithm.

One of the critical steps in the $\sqrt{\text{AO}}$ and the AO algorithms
is the determination of the map $\tau\mapsto\busf{W}(\tau)$.
Here $\busf{W}(\tau)$ is time dependent skew-symmetric matrix of real numbers that is related to the rigid body's angular velocity.
In the AO-algorithm $\busf{W}$ is determined by numerically integrating the equation (\cite[\S3.12]{rahaman2019})
\begin{equation}
\overline{\usf{W}}'(\tau)=\textrm{skew part of }\usf{P}(\tau).\label{equ:skew}
\end{equation}
Here $\usf{P}(\tau)$ is a square matrix of real numbers that is to
be computed from the accelerometers' data, relative locations,
and orientations. Due to numerical integration any bias type errors
in $\usf{P}$ will give rise to errors in $\busf{W}$ that grow with
time. In the $\sqrt{\text{AO}}$-algorithm we alternatively determine $\busf{W}$ by taking the square-root of the equation
\begin{equation}
\overline{\usf{W}}(\tau)\overline{\usf{W}}(\tau)=\textrm{symmetric part of }\usf{P}(\tau).\label{equ:sym-1}
\end{equation}
We derive \eqref{equ:sym-1} in \ref{subsec:square}. Due to the elimination of the numerical integration step associated with the solution of $\eqref{equ:skew}$, the $\sqrt{\textrm{AO}}$-algorithm
gives much better persistent accuracy over time when applied to the data containing bias type errors, compared to the AO-algorithm.

In $\S$\ref{Sec:Math} we present the mathematics and mechanics of
rigid body motion from \cite[\S2]{rahaman2019} that is needed for
the development of the $\sqrt{\text{AO}}$-algorithm.
In $\S$\ref{subsec:Review-of-the} we review the AO-algorithm as preparation for the development of the $\sqrt{\text{AO}}$-algorithm.
In $\S$\ref{Sec:PM} we detail the $\sqrt{\textrm{AO}}$-algorithm and present a procedure for taking the square root of $\eqref{equ:sym-1}$.
In $\S$\ref{sec:In-silico-validation} we check the validity and
robustness of the $\sqrt{\text{AO}}$-algorithm. We do so by feeding
in virtual accelerometer data, to which differing amounts of bias
and noise type errors have been added, to both the $\sqrt{\text{AO}}$
and $\text{AO}$ algorithms and comparing their predictions. Using
those predictions in $\S$\ref{sec:Results-and-Discussion} we show
that the $\sqrt{\textrm{AO}}$-algorithm is less sensitive to bias
type errors than the AO-algorithm. We make a few concluding remarks
in $\S$\ref{Sec:Con}.

\section{Preliminary mathematics and kinematics of rigid body motion}

\label{Sec:Math}

In this section we briefly recapitulate the mathematics and kinematics
of rigid body motion from \cite[\S2]{rahaman2019} that are needed for
the development of the proposed $\sqrt{\textrm{AO}}$-algorithm.

\subsection{Notation}
\label{subsec:notation}

\begin{figure}[ht]
\centering{}
\includegraphics[width=\textwidth]{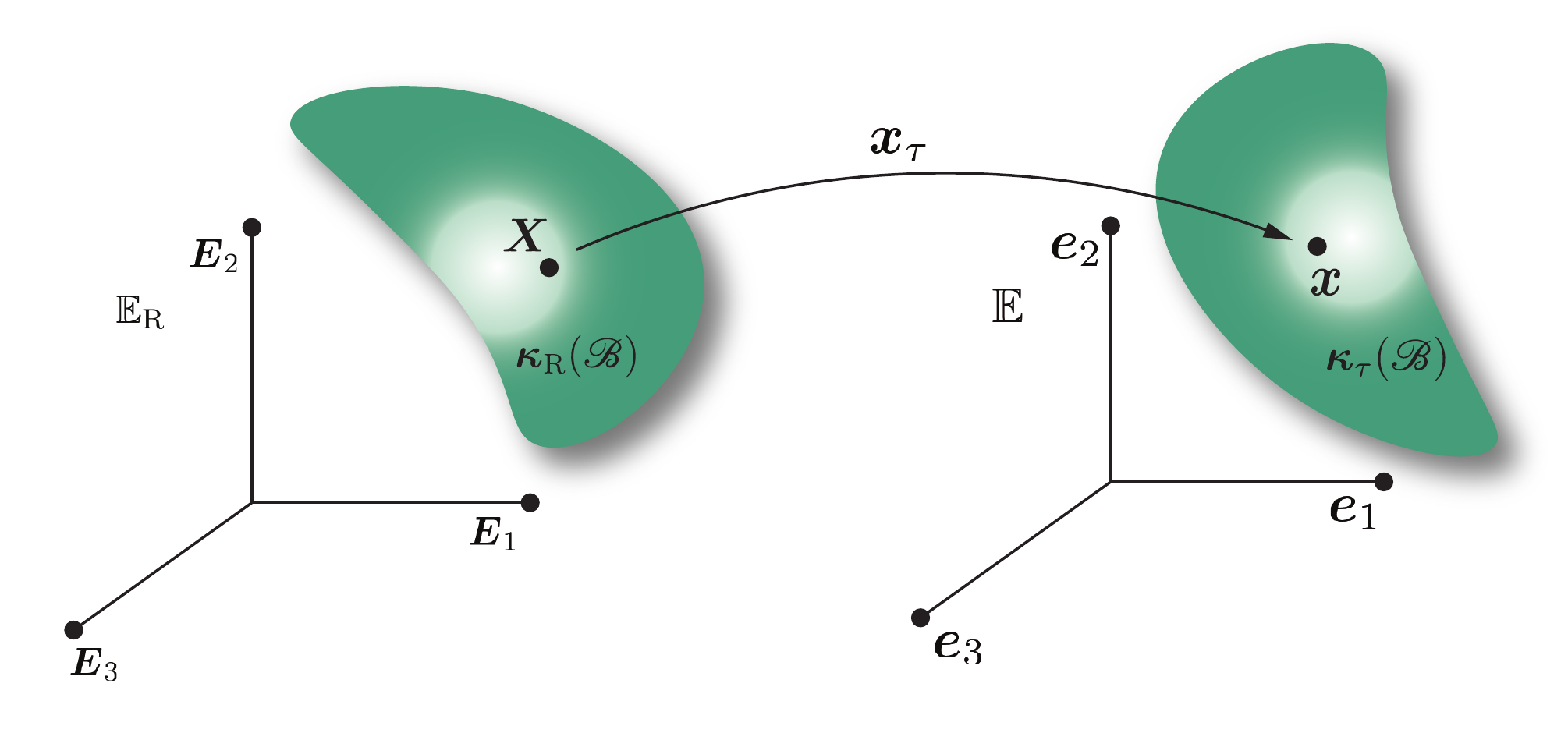}
\caption{Some mathematical quantities used in the description of motion. Illustration of the reference Euclidean vector space $\mathbb{E}_{\rm R}$, reference body $\u{\kappa}_{{\rm R}}(\mathcal{B})$, a material particle $\boldsymbol{X}$, the deformation map $\boldsymbol{x}_{\tau}$, current body $\u{\kappa}_{\tau}\pr{\mathcal{B}}$, the (physical) Euclidean vector space $\mathbb{E}$, and the location of the material particle $\u{X}$ in $\mathbb{E}$, i.e., the material particle $\boldsymbol{X}$'s spatial position vector $\boldsymbol{x}$. See $\S$\ref{subsec:notation} for details.}
\label{fig:motion}
\end{figure}

Let $\mathbb{E}$ be a finite dimensional, oriented, Hilbert space,
i.e., a Euclidean vector space. The Euclidean point space $\mathcal{E}$
has $\mathbb{E}$ as its associated vector space. Let $o\in\mathcal{E}$
be $\mathcal{E}$'s origin. The spaces $\mathbb{E}$ and $\mathcal{E}$
are related to each other such that for any point $x\in\mathcal{E}$
there exists a vector $\u{x}\in\m{E}$ such that $o+\boldsymbol{x}=x$.
The topological space $\mathcal{B}$ serves as our model for a rigid
body that executes its motion in $\mathcal{E}$. For that reason, we
refer to $\mathbb{E}$ and $\mathcal{E}$ as the physical Euclidean
vector space and point space, respectively. The
spaces $\mathbb{E}_{{\rm R}}$ and $\mathcal{E}_{\rm R}$ are another
pair of Euclidean vector and point spaces, respectively, that are
related to each other in the same way that $\mathbb{E}$ and $\mathcal{E}$
are related to each other. We refer to $\mathbb{E}_{{\rm R}}$ and $\mathcal{E}_{\rm R}$
as the reference Euclidean vector and point spaces, respectively.
The spaces $\mathbb{E}$, $\mathcal{E}$, $\mathbb{E}_{{\rm R}}$, and $\mathcal{E}_{\rm R}$
have the same dimension, which we denote as $n_{\rm sd}$. The dimension
of $\mathcal{B}$ is less than or equal to $\nsd$. We
call a select continuous, injective map from $\mathcal{B}$ into $\mathbb{E}_{{\rm R}}$
the reference configuration and denote it as $\boldsymbol{\kappa}_{{\rm R}}$.
The elements of $\mathcal{B}$ are called material particles. We call $\u{X}=\boldsymbol{\kappa}_{{\rm R}}(\mathcal{X})$,
where $\mathcal{X}\in\mathcal{B}$, the reference position vector of
the material particle $\mathcal{X}$, and we call
the set $\u{\kappa}_{{\rm R}}(\mathcal{B})=\set{\u{\kappa}_{{\rm R}}(\mathcal{X})\in\mathbb{E}_{{\rm R}}}{\mathcal{X}\in\mathcal{B}}$
the reference body (see Fig.~\ref{fig:motion}).
When we refer to $\u{X}$ as a material particle we in fact mean the
material particle $\boldsymbol{\kappa}_{{\rm R}}^{-1}(\boldsymbol{X})\in\mathcal{B}$.
We model time as a one-dimensional normed vector
space $\m{T}$ and denote a typical element in it
as $\u{\tau}=\tau\u{s}$, where $\tau\in\m{R}$
and $\u{s}$ is a fixed vector in $\m{T}$ of unit norm. We model the
rigid body's motion using the one-parameter family of maps $\boldsymbol{\u{x}}_{{\rm \tau}}:\mathcal{\mathbb{\mathbb{\mathbb{E}_{{\rm R}}}}}\rightarrow\mathbb{E}$
(see Fig.~\ref{fig:motion}). We call $\u{x}_{\tau}$
the deformation map and $\u{x}=\u{x}_{\tau}(\u{X})$ the material particle $\u{X}$'s
position vector at the time instance $\u{\tau}$. The set $\u{\kappa}_{\tau}\pr{\mathcal{B}}$=$\set{\u{x}_{\tau}(\u{X})\in\mathbb{E}}{\u{X}\in\u{\kappa}_{\rm R}\pr{\mathcal{B}}}$ (see Fig.~\ref{fig:motion}) is called the current body.

\subsection{Components}

The sets $\pr{\u{E}_{i}}_{i\in\mathcal{I}}$ and $\pr{\u{e}_{i}}_{i\in\mathcal{I}}$, where $\mathcal{I}=\pr{1,\ldots,\nsd}$, are orthonormal sets of basis vectors for $\mathbb{E}_{{\rm R}}$ and $\mathbb{E}$, respectively.
By orthonormal we mean that the inner product between $\u{E}_i$ and $\u{E}_j$,
or $\u{e}_{i}$ and $\u{e}_j$, where $i,j\in\mathcal{I}$, equals $\delta_{ij}$,
the Kronecker delta symbol, which equals unity iff $i=j$ and zero otherwise.
We call $X_{i}$ the component of $\u{X}$ w.r.t. $\u{E}_{i}$ iff $X_i=\u{X}\cdot\u{E}_i$, where the dot denotes
the inner product in $\mathbb{E}_{\rm R}$. The dot in other expressions
is to be similarly interpreted noting the space to which the vectors belong. We call the ordered set $\pr{X_{i}}_{i\in\mathcal{I}}$ the component form of $\u{X}$ w.r.t. $\pr{\u{E}_{i}}_{i\in\mathcal{I}}$ and denote it as $\usf{X}$
or $\mathcal{M}\!\u{X}$. We denote the space of all $m\times n$ real
matrices, where $m,n\in\mathbb{N}$, $\mathcal{M}_{m,n}(\mathbb{R})$;
here $\mathbb{N}$ and $\mathbb{R}$ denote the set of natural numbers
and the space of real numbers, respectively. Thus, $\usf{X}\in\mathcal{M}_{n_{{\rm sd}},1}(\mathbb{R})$.
We access the $i^{\rm th}$ component, where $i\in\mathcal{I}$, of $\usf{X}$,
which of course is $X_i$, as $\pr{\usf{X}}_{i}$. Similarly, we denote
the component of $\u{x}$ w.r.t. $\u{e}_{i}$ as $x_{i}$ and call $\usf{x}=\pr{x_{i}}_{i\in\mathcal{I}}\in\mathcal{M}_{n_{{\rm sd}},1}(\mathbb{R})$
the component form of $\u{x}$ w.r.t. $\pr{\u{e}_{i}}_{i\in\mathcal{I}}$.

Say $\mathbb{W}$ and $\mathbb{U}$ are two arbitrary, oriented, finite dimensional Hilbert spaces; for instance, they can
be $\mathbb{E}_{\rm R}$ and $\mathbb{E}$. We denote the space of all
linear maps (transformations/operators) from $\mathbb{W}$ to $\mathbb{U}$
as $\ensuremath{\mathcal{L}}(\mathbb{W},\mathbb{U})$\footnote{In our previous paper \cite{rahaman2019}, we denoted the set of bounded
linear operators from $\mathbb{U}$ to $\mathbb{W}$ as $B(\mathbb{U},\mathbb{W})$. As a linear operator on a finite dimensional normed space is automatically bounded, here we use $\ensuremath{\mathcal{L}}(\mathbb{U},\mathbb{W})$ instead of $B(\mathbb{U},\mathbb{W})$ to denote the set of all linear
operators from $\mathbb{U}$ to $\mathbb{W}$.}.
We denote the norm of a vector $\u{w}_{1}$ in $\mathbb{W}$ that
is induced by $\mathbb{W}$'s inner product, i.e., $(\u{w}_{1}\cdot\u{w}_{1})^{1/2}$, as $\lVert\u{w}_{1}\rVert$. For $\u{u}_{1}\in\mathbb{U}$, the expression $\u{u}_{1}\otimes\u{w}_{1}$
denotes the linear map from $\mathbb{W}$ to $\mathbb{U}$
defined as
\begin{equation}
\pr{\u{u}_{1}\otimes\u{w}_{1}}\u{w}_{2}=\u{u}_{1}\pr{\u{w}_{1}\cdot\u{w}_{2}},
\end{equation}
where $\u{w}_{2}\in\mathbb{W}$. If the sets $\pr{\u{u}_i}_{i\in\mathcal{I}}$
and $\pr{\u{w}_i}_{i\in\mathcal{I}}$ provide bases for $\mathbb{U}$
and $\mathbb{W}$, respectively, then it can be shown that $\pr{\pr{\u{u}_{i}\otimes\u{w}_j}_{j\in\mathcal{I}}}_{i\in\mathcal{I}}$,
which we will henceforth abbreviate as $\pr{\u{u}_{i}\otimes\u{w}_j}_{i,j\in\mathcal{I}}$, provides a basis for $\ensuremath{\mathcal{L}}(\mathbb{W},\mathbb{U})$. The number $T_{ij}$, where $i,j\in\mathcal{I}$, is called the component of $\u{T}\in\mathcal{L}(\mathbb{W},\mathbb{U})$
w.r.t. $\u{u}_{i}\otimes\u{w}_j$ iff $T_{ij}=\u{u}_{i}\cdot\pr{\u{T}\u{w}_j}$. We call the nested ordered set $\pr{T_{ij}}_{i,j\in\mathcal{I}}$ the component
form of $\u{T}$ w.r.t. $\pr{\u{u}_{i}\otimes\u{w}_j}_{i,j\in\mathcal{I}}$,
and denote it as $\mathcal{M}\u{T}$, or, when possible, briefly as $\usf{T}$.
We sometimes access the $i^{\rm th}$, $j^{\rm th}$ component of $\usf{T}$,
where $i,j\in\mathcal{I}$, as $\pr{\usf{T}}_{ij}$.

From here on, unless otherwise specified, we will be following the
Einstein summation convention. As per this convention a repeated
index in a term will imply a sum over that term with the repeated index taking values in $\mathcal{I}$. For example, the expression $X_{i}\u{E}_{i}$ represents the sum $\sum_{i\in\mathcal{I}}X_{i}\u{E}_{i}$.
And an unrepeated index in a term will signify a set of $\nsd$ terms. For example, the term $\u{E}_{i}$ represents the set $\set{\u{E}_{i}}{i\in\mathcal{I}}$.

\subsection{Velocities and Accelerations}
\label{subsec:Velocities-and-Accelerations}

For the case of rigid body motion $\u{x}_{\tau}$ takes the form
\begin{equation}
\boldsymbol{x}_{\tau}(\boldsymbol{X})=\boldsymbol{Q}_{\tau}\boldsymbol{X}+\boldsymbol{c}(\boldsymbol{\tau}),
\label{equ:position}
\end{equation}
where $\u{Q}_{\tau}$ is a proper (orientation preserving), linear isometry from $\mathbb{E}_{{\rm R}}$ into $\mathbb{E}$ and $\u{c}(\u{\tau})=c_{i}(\tau)\u{e}_{i}$, where $c_{i}$ belongs to the space of twice continuously differentiable real valued functions
over $\mathbb{R}$, i.e., to $C^{2}\pr{\mathbb{R},\mathbb{R}}$.
The operator $\u{Q}_{\tau}$ can be written as $Q_{ij}(\tau)\u{e}_{i}\otimes\u{E}_{j}$, where $Q_{ij}\in C^{2}\pr{\mathbb{R},\mathbb{R}}$ and satisfy $Q_{ki}(\tau) Q_{kj}(\tau)=\delta_{ij}$ for all $\tau\in\mathbb{R}$.
We abbreviate $\pr{Q_{ij}(\tau)}_{i,j\in\mathcal{I}}\in\mathcal{M}_{\nsd,\nsd}(\mathbb{R})$, $\pr{c_{i}(\tau)}_{i\in\mathcal{I}}\in\mathcal{M}_{\nsd,1}(\mathbb{R})$, and $(\delta_{ij})_{i,j\in\mathcal{I}}\in\mathcal{M}_{\nsd,\nsd}(\mathbb{R})$ as $\usf{Q}(\tau)$, $\usf{c}(\tau)$, and $\usf{I}$, respectively.
The component or non-dimensional form of \eqref{equ:position} is
$\eqref{eq:NondimensionalRotationDeformationMapping}$.
Since $\u{Q}_{\tau}$ is a proper isometry, it follows that $\usf{Q}(\tau)$,
which we refer to as the rotation matrix, belongs to the special orthogonal
group $SO(\nsd)$. As a consequence of belonging to $SO(\nsd)$ the
matrix $\usf{Q}(\tau)$ satisfies the equations

\begin{subequations}
\begin{align}
\tpsb{\usf{Q}}(\tau)\,\usf{Q}(\tau)&=\usf{I},\label{equ:QQ:1}
\intertext{and}
\usf{Q}(\tau)\,\tpsb{\usf{Q}}(\tau)&=\usf{I},\label{equ:QQ:2}
\end{align}
\label{eq:QQ}
\end{subequations}
where $\tpsb{\usf{Q}}(\tau)$ is the transpose of $\usf{Q}(\tau)$, i.e., $\tpsb{\usf{Q}}(\tau)=\tpsb{\pr{\usf{Q}(\tau)}}$.

We call $\mathcal{L}\pr{\mathbb{T},\mathbb{E}}$ the physical velocity
vector space and denote it as $\mathbb{V}$. It can be shown that the
set $\pr{\u{v}_{i}}_{i\in\mathcal{I}}$, where $\u{v}_{i}\in\mathbb{V}$
and are defined such that $\u{v}_{i}\u{\tau}=\tau\u{e}_i$, provides
an orthonormal basis for $\mathbb{V}$. The velocity of a material
particle $\u{X}$ executing its motion in $\mathbb{E}$ lies in $\mathbb{V}$.
The velocity of the material particle $\u{X}$ at the instant $\u{\tau}$, which we denote as $\u{V}_{\tau}(\u{X})$, equals the value of the Fr\'echet derivative\footnote{For the definition of Fr\'echet derivative in the context of the current work see \cite[\S2.1]{rahaman2019}} of the map $\mathbb{T}\ni\u{\tau}\mapsto\u{x}_{\u{X}}(\u{\tau})\in\mathbb{E}$,
where $\u{x}_{\u{X}}(\u{\tau})=\u{x}_{\tau}(\u{X})$, at the time instance $\u{\tau}$. Thus, it follows from \eqref{equ:position} that
\begin{equation}
\u{V}_{\tau}(\u{X})=\u{L}_{\tau}\u{X}+\u{c}'(\u{\tau}),
\label{equ:vel1}
\end{equation}
where $\u{L}_{\tau}:=Q'_{ij}(\tau)\u{v}_{i}\otimes\u{E}_j$ and $\u{c}'(\u{\tau}):=c_{i}'(\tau)\u{v}_{i}$, and $Q'_{ij}$ and $c_{i}'$ are the derivatives of $Q_{ij}$ and $c_{i}$, respectively. We abbreviate $\pr{Q_{ij}'(\tau)}_{i,j\in\mathcal{I}}\in\mathcal{M}_{\nsd,\nsd}(\mathbb{R})$ and $\pr{c_{i}'(\tau)}_{i\in\mathcal{I}}\in\mathcal{M}_{\nsd,1}(\mathbb{R})$
as $\usf{Q}'(\tau)$ and $\usf{c}'(\tau)$, respectively.
Using \eqref{equ:vel1} and \eqref{equ:position} it can be shown that the
velocity at the time instance $\u{\tau}$ of the material particle occupying the spatial position $\u{x}\in\mathbb{E}$ at the time instance $\u{\tau}$ is $\u{W}_{\tau}\pr{\boldsymbol{x}-\u{c}(\u{\tau})}+\u{c}'(\u{\tau})$,
where the linear map $\u{W}_{\tau}:\mathbb{E}\to\mathbb{V}$ is defined
by the formula
\begin{equation}
\u{W}_{\tau}\u{x}=\u{L}_{\tau}\u{Q}_{\tau}^{*}\u{x},
\label{equ:wde}
\end{equation}
for all $\u{x}\in\m{E}$. The operator $\u{Q}_{\tau}^{*}$ is the Hilbert-adjoint of $\u{Q}_{\tau}$ and is equal to $Q_{ji}(\tau)\u{E}_{i}\otimes\u{e}_{j}$. Let the component form of $\u{W}_{\tau}$ w.r.t. $\pr{\u{v}_i\otimes\u{e}_j}_{i,j\in\mathcal{I}}$ be $\pr{W_{ij}(\tau)}_{i,j\in\mathcal{I}}\in\mathcal{M}_{\nsd,\nsd}(\mathbb{R})$, which we abbreviate as $\usf{W}(\tau)$. It follows from \eqref{equ:wde}
that $W_{ij}(\tau)=Q'_{ik}(\tau)Q_{jk}(\tau)$, or equivalently,
\begin{equation}
\usf{W}(\tau)=\usf{Q}'(\tau)\tpsb{\usf{Q}}(\tau).
\label{equ:wdeNDF}
\end{equation}

We call $\mathcal{L}\pr{\mathbb{T},\mathbb{V}}$ the physical acceleration
vector space and denote it as $\mathbb{A}$. It can be shown that the
set $\pr{\u{a}_{i}}_{i\in\mathcal{I}}$, where $\u{a}_{i}\in\mathbb{A}$
and are defined such that $\u{a}_{i}\u{\tau}=\tau\u{v}_i$, provides
an orthonormal basis for $\mathbb{A}$. The acceleration of a material
particle $\u{X}$ executing its motion in $\mathbb{E}$ lies in $\mathbb{A}$.
The acceleration of $\u{X}$ at the time instance $\u{\tau}$
equals the value of the Fr\'echet derivative of the map $\mathbb{T}\ni\u{\tau}\mapsto\u{V}_{\u{X}}(\u{\tau})\in\mathbb{V}$, where $\u{V}_{\u{X}}(\u{\tau})=\u{V}_{\tau}(\u{X})$, at the
time instance $\u{\tau}$. Thus, it follows from \eqref{equ:vel1} that
\begin{equation}
\u{A}_{\tau}(\u{X})={\color{red}}\u{M}_{\tau}\u{X}+\u{c}''(\u{\tau}),
\end{equation}
where the map $\u{M}_{\tau}:\m{E}_{\rm R}\to\m{A}$ is defined by the
equation
\begin{equation}
\u{M}_{\tau}:=Q''_{ij}(\tau)\u{a}_{i}\otimes\u{E}_j
\label{eq:MtauDef}
\end{equation}
and $\u{c}''(\u{\tau}):=c_{i}''(\tau)\u{a}_{i}$, where $Q''_{ij}$ and $c''_{i}$
are the derivatives of $Q'_{ij}$ and $c'_{i}$, respectively.
Let $A_{\tau i}(\boldsymbol{X})$ be the component of $\u{A}_{\tau}(\u{X})$ w.r.t. $\u{a}_{i}$. We abbreviate the ordered sets $\pr{A_{\tau i}(\boldsymbol{X})}_{i\in\mathcal{I}}\in\mathcal{M}_{n_{{\rm sd}},1}(\mathbb{R})$, $\pr{Q_{ij}''(\tau)}_{i,j\in\mathcal{I}}\in\mathcal{M}_{\nsd,\nsd}(\mathbb{R})$, and $\pr{c_{i}''(\tau)}_{i\in\mathcal{I}}\in\mathcal{M}_{\nsd,1}(\mathbb{R})$ as $\usf{A}_{\tau}(\usf{X})$, $\usf{Q}''(\tau)$, and $\usf{c}''(\tau)$, respectively.

We will predominantly be presenting the ensuing results in component form. The component form can be converted into physical or dimensional form. Therefore, from here on we will be often omit explicitly using the qualification ``is the component form of'' when referring to the component form of a physical quantity. For example, instead of saying ``$\usf{A}_{\tau}(\usf{X})$ as the component form of the acceleration of the material particle $\u{X}$ at the time instance $\u{\tau}$'', we will often write ``$\usf{A}_{\tau}(\usf{X})$ is the acceleration of the material particle $\usf{X}$ at the time instance $\u{\tau}$''.
The acceleration $\usf{A}_{\tau}(\usf{X})$ can be interpreted as the value of the (non-dimensional) acceleration field $\usf{A}_{\tau}:B_{\rm R}(\mathcal{B})\to\mathbb{R}^3$, where we call $B_{\rm R}(\mathcal{B}):=\set{(X_1,X_2,X_3)\in\mathbb{R}^3}{X_i\u{E}_i\in\u{\kappa}_{{\rm R}}(\mathcal{B})}$ the non-dimensional reference body.

\section{Review of the AO-algorithm}
\label{subsec:Review-of-the}
Let $\overline{\u{Q}}_{\tau}:\m{A}\to\m{E}_{\rm R}$ be defined by the equation
\begin{equation}
\overline{\u{Q}}_{\tau}=Q_{ji}(\tau)\u{E}_i\otimes\u{a}_{j},
\label{eq:OverbarQtauDef}
\end{equation}
then we call the map $\overline{\u{A}}_{\tau}:\u{\kappa}_{{\rm R}}(\mathcal{B})\to\m{E}_{\rm R}$ defined by the equation
\begin{equation}
\overline{\boldsymbol{A}}_{\tau}(\u{X})=\overline{\boldsymbol{Q}}_{\tau}\boldsymbol{A}_{\tau}(\u{X})\label{eq:barAtauDef}
\end{equation}
the ``Pseudo-acceleration field''.
Say $\overline{A}_{\tau i}(\boldsymbol{X})$ is the component of $\overline{\u{A}}_{\tau}(\u{X})$
w.r.t. $\u{E}_{i}$, then we abbreviate $\pr{\overline{A}_{\tau i}(\boldsymbol{X})}_{i\in\mathcal{I}}\in\mathcal{M}_{n_{{\rm sd}},1}(\mathbb{R})$, the component form of $\overline{\u{A}}_{\tau}(\u{X})$ w.r.t. $\u{E}_i$, as $\busf{A}_{\tau}(\usf{X})$.
From the definitions of the pseudo acceleration field $\overline{\u{A}}_{\tau}$ \eqref{eq:barAtauDef}, and $\busf{A}_{\tau}(\usf{X})$, and the definitions of $\usf{A}_{\tau}(\usf{X})$, and $\usf{Q}(\tau)$, which are given
in $\S$\ref{subsec:Velocities-and-Accelerations}, it follows that
\begin{align}
\usf{A}_{\tau}(\usf{X})=\usf{Q}(\tau)\busf{A}_{\tau}(\usf{X}).\label{equ:acce}
\end{align}
In \cite[\S2.1.1]{rahaman2019} it was shown that \begin{equation}
\busf{A}_{\tau}(\usf{X})=\usf{P}(\tau)\usf{X}+\usf{q}(\tau),
\label{equ:psuedo}
\end{equation}
where
\begin{equation}
\usf{P}(\tau)=\usf{Q}^{\sf T}(\tau)\,\usf{Q}''(\tau)\label{equ:pde}
\end{equation}
is the component form of
the linear map $\u{P}_{\tau}:=\overline{\u{Q}}_{\tau}\circ\u{M}{}_{\tau}$
w.r.t. $\left(\u{E}_{i}\otimes\u{E}_{j}\right)_{i,j\in\mathcal{I}}$,
and $\usf{q}(\tau)\in\mathcal{M}_{n_{{\rm sd}},1}(\mathbb{R})$ is the component form of
\begin{equation}
\u{q}(\u{\tau}):=\overline{\u{Q}}_{\tau}\u{c}''(\u{\tau})
\end{equation}
w.r.t. $\pr{\u{E}_{i}}_{i\in\mathcal{I}}$. Thus, the acceleration field $\usf{A}_{\tau}$ is taken to be fully determined once $\usf{Q}(\tau)$, $\usf{P}(\tau)$, and $\usf{q}(\tau)$ have been computed.

Both the AO and the $\sqrt{\text{AO}}$ algorithms can be described
as consisting of three primary steps. The AO-algorithm's three steps can briefly be described as follows:
\begin{enumerate}[label=\textit{AO-Step~\arabic*},leftmargin=1.7cm]
  \item Compute (time discrete versions of) the maps $\tau\mapsto\usf{P}(\tau)$
  and $\tau\mapsto\usf{q}(\tau)$ using the measurements and the geometry
  of the arrangement of the four tri-axial accelerometers.\label{AOstep1}

  \item Compute the map $\tau\mapsto\busf{W}(\tau)$, where
  \begin{equation}
  \busf{W}(\tau):=\tpsb{\usf{Q}}(\tau)\usf{W}(\tau)\usf{Q}(\tau)
  \label{eq:busfW},
  \end{equation}
  using the map $\usf{P}$ computed in \ref{AOstep1} and numerical integrating \eqref{equ:skew}.
  From Lemma \ref{lemma:skew} we have that the matrix $\busf{W}(\tau)$
  belongs to the space of $\nsd\times\nsd$ real skew-symmetric matrices,
  which we denote as $\mathfrak{so}(\mathbb{R},\nsd)$.\label{AOstep2}

  \item Compute the map $\tau\mapsto\usf{Q}(\tau)$ using the $\busf{W}$ map
  computed in \ref{AOstep2} and numerically integrating the equation
  \begin{equation}
  \usf{Q}'(\tau)=\usf{Q}(\tau)\busf{W}(\tau).
  \label{equ:rotation}
  \end{equation}
  Equation $\eqref{equ:rotation}$ is from \cite{rahaman2019}, where it appears as equation $3.14$.\label{AOstep3}
\end{enumerate}

Step one of the $\sqrt{\text{AO}}$-algorithm has two sub-steps: the \textit{predictor} step and the $\textit{corrector}$ step. The predictor
step is the same as \ref{AOstep1} of the AO-algorithm. The
corrector step is necessary for carrying out step two of the $\sqrt{\text{AO}}$-algorithm.
In step two of the $\sqrt{\text{AO}}$-algorithm instead of obtaining $\busf{W}$
from $\eqref{equ:skew}$, as is done in the AO-algorithm, we obtain it from \eqref{equ:sym-1}.
More precisely, in the $\sqrt{\text{AO}}$-algorithm $\busf{W}(\tau)$
is obtained as the square root of the symmetric part of $\usf{P}(\tau)$.
We use $\text{sym}\pr{\usf{P}(\tau)}$ to denote the symmetric part of $\usf{P}(\tau)$.
The derivation of \eqref{equ:sym-1} is presented in \ref{subsec:square}.
A procedure for determining $\overline{\usf{W}}(\tau)$ as the square
root of $\text{sym}\pr{\usf{P}(\tau)}$, i.e., for solving \eqref{equ:sym-1}
for $\busf{W}(\tau)$ with given $\usf{P}(\tau)$, is presented in $\S$\ref{step2}.
The goal of step three of the $\sqrt{\text{AO}}$-algorithm is to compute $\usf{Q}$ using the $\busf{W}$ computed in step two. It involves using a slightly modified version of the numerical integration scheme described by
equations $3.15$, $3.16$, and $3.17$ in \cite[\S3.2]{rahaman2019} to solve $\eqref{equ:rotation}$. We discuss it in $\S$\ref{step3}.

\section{The $\sqrt{\text{AO}}$-algorithm}
\label{Sec:PM}

As we mentioned in $\S$\ref{subsec:Review-of-the} the $\sqrt{\text{AO}}$ algorithm consists of three primary steps. Those steps are as follows:
\begin{enumerate}[label=Step~\arabic*,leftmargin=1.2cm]
\item Compute (time discrete versions of) the maps $\tau\mapsto\usf{P}(\tau)$ and $\tau\mapsto\usf{q}(\tau)$ using the measurements and the geometry of the arrangement of the four tri-axial accelerometers (see $\S\ref{step1}$ for details).\label{enu:sqrtAO-Step1}
\item Use \eqref{equ:sym-1} and the $\usf{P}$ map obtained from \ref{enu:sqrtAO-Step1} to solve for $\tau\mapsto\busf{W}(\tau)$. That is, for each $\tau$ in a discrete sequence of time instances, compute $\busf{W}(\tau)$ as the square root of the symmetric part of $\usf{P}(\tau)$ (for details see $\S\ref{step2}$).\label{enu:sqrtAO-Step2}
\item Compute (a time discrete version of) the map $\tau\mapsto\usf{Q}(\tau)$ using the $\busf{W}$ map computed in \ref{enu:sqrtAO-Step2} and numerically integrating $\eqref{equ:rotation}$ (details in $\S$\ref{step3}).\label{enu:sqrtAO-Step3}
\end{enumerate}

\subsection{$\sqrt{\text{AO}}$-algorithm, \ref{enu:sqrtAO-Step1} of $3$}
\label{step1}

In $\S3.1$ of \cite{rahaman2019} a method was presented to estimate $\usf{P}(\tau)$ and $\usf{q}(\tau)$ from the accelerometer measurements corresponding to the time instance $\tau$. Applying that method for each $\tau$ in a discrete time sequence yields a numerical approximation for the maps $\tau\mapsto\usf{P}(\tau)$ and $\tau\mapsto\usf{q}(\tau)$.
We present here an augmented version of that method for computing
similar numerical approximations. The
primary difference between our method and that presented in \cite{rahaman2019}
is that the estimate for $\usf{P}(\tau)$ yielded by our method is
certain to retain some of the mathematical properties that are expected
of $\usf{P}(\tau)$ based on our theoretical analysis.
Specifically, it follows from Lemmas ~\ref{lemma:negative} and \ref{lemma:form} that $\textrm{sym}\pr{\usf{P}(\tau)}$ is a negative semidefinite matrix
with its negative eigenvalues, if any, being of even algebraic multiplicities.
These mathematical properties of $\usf{P}(\tau)$ are critical for carrying out  \ref{enu:sqrtAO-Step2} of the $\sqrt{\text{AO}}$-algorithm.
We found that experimental noise and errors can cause the estimate
for $\usf{P}(\tau)$ provided by the method presented in \cite{rahaman2019}
to lose the aforementioned mathematical properties.
Our method, on the contrary, ensures that the symmetric part of the
estimated $\usf{P}(\tau)$ is negative semidefinite and that its negative
eigenvalues, when they exist, are of even algebraic multiplicities.
Once $\usf{P}(\tau)$ is estimated, our method to estimate $\usf{q}(\tau)$
is exactly the same as that in \cite{rahaman2019}.
We review it in \ref{subsec:Estimating-1}.

\subsubsection{Estimating $\usf{P}(\tau)$\label{subsec:Estimating}}

Our method for estimating $\usf{P}(\tau)$ can be described as consisting
of two steps: a $\textit{predictor step}$ and a $\textit{corrector step}$.
In the predictor step we use the method presented in \cite[\S3.1]{rahaman2019}
for estimating $\usf{P}(\tau)$ to compute a prediction for $\usf{P}(\tau)$.
We denote this prediction as $\usf{P}(\tau)^{\rm p}$.
In the corrector step we estimate $\usf{P}(\tau)$ as the sum of $\usf{P}(\tau)^{\rm p}$ and a correction term, which we construct using $\usf{P}(\tau)^{\rm p}$.
The correction term is constructed such that the estimated $\usf{P}(\tau)$
is as close as possible to $\usf{P}(\tau)^{\rm p}$ under the constraint
that the estimated $\usf{P}(\tau)$'s symmetric part is negative semidefinite
and its negative eigenvalues (if they exist) are of even algebraic multiplicities.

\paragraph{Predictor step}

\begin{figure}[ht!]
\centering{}\includegraphics[width=\textwidth]{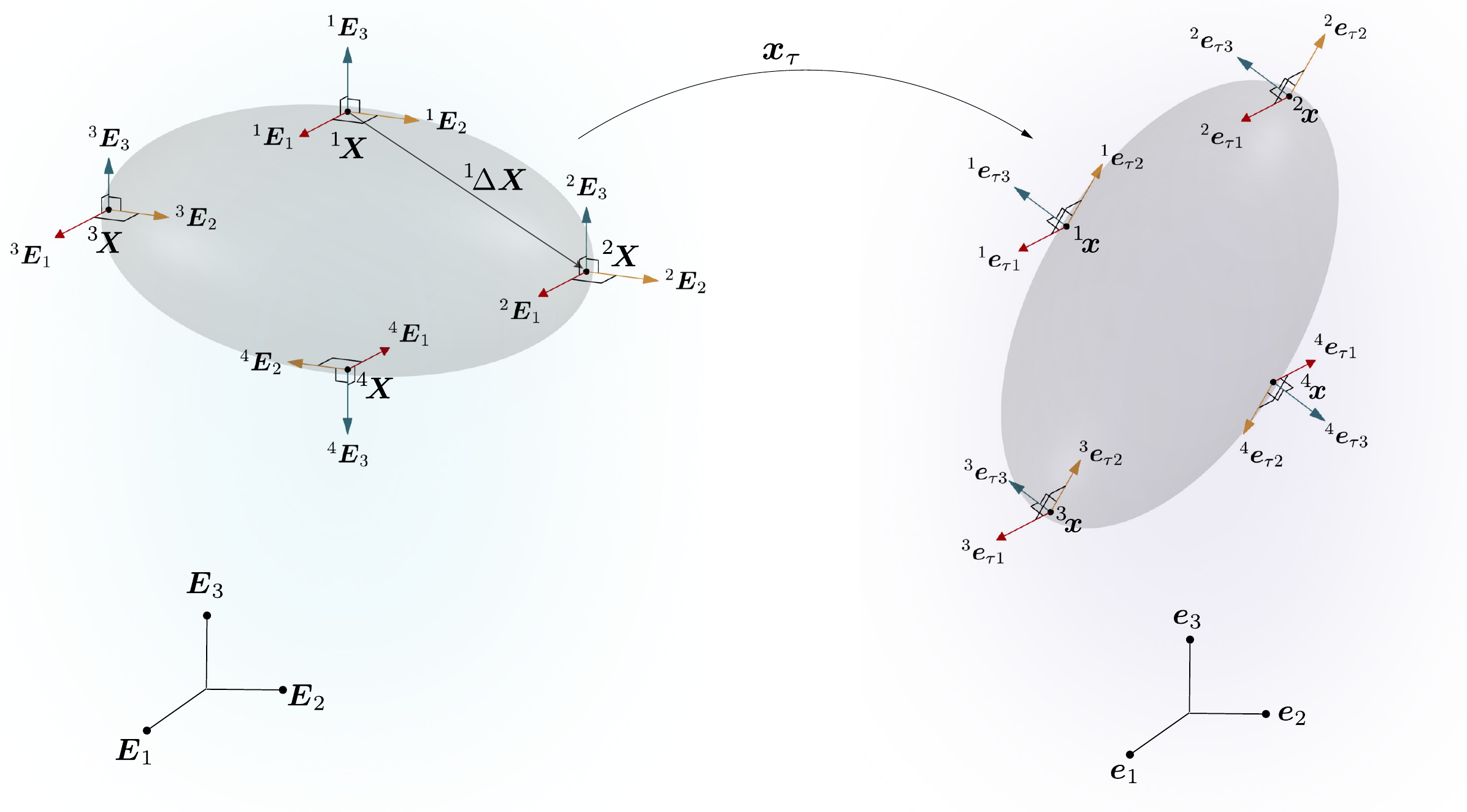}\caption{Schematic of the  locations and orientations of four tri-axial accelerometers (left) and their motion (right) (modified from \cite{rahaman2019}, copyright 2020, Elsevier).}
\label{fig:EllipsoidAccDir}
\end{figure}

Say the four tri-axial accelerometers are attached to the rigid body $\c{B}$
at the material particles $\pr{\lsc{\mathcal{X}}}_{\ell\in\mathcal{J}}$, where $\mathcal{J}:=(1,\ldots,4)$, and let the position vectors of
those particles in $\m{E}_{\rm R}$, respectively, be $\pr{\lsc{\u{X}}}_{\ell\in\mathcal{J}}$ (see Fig.~\ref{fig:EllipsoidAccDir}).
A tri-axial accelerometer is capable of measuring the components of
its acceleration in three mutually perpendicular directions.
We refer to those directions as the accelerometer's measurement directions.
The measurement directions are usually marked on the accelerometer package by the manufacturer as arrows that are labeled $x$, $y$, and $z$.
As $\c{B}$ moves in $\m{E}$, the attached accelerometers move with
it, and, therefore, the measurement directions (in $\mathbb{E}$) can
change with time.
For an accelerometer $\ell$, where $\ell\in\mathcal{J}$, we denote its time varying measurement directions in $\mathbb{E}$ using the orthonormal set $\pr{\lsc{\u{e}}_{\tau i}}_{i\in\mathcal{I}}$.
Assuming that the accelerometers remain rigidly attached to $\mathcal{B}$,
i.e., their positions and orientations w.r.t. $\mathcal{B}$ do not
change as $\mathcal{B}$ moves in $\m{E}$, it can be shown that $\u{Q}^*_{\tau}\,\lsc{\u{e}}_{\tau i}$, where $\ell\in\mathcal{J}, i\in\mathcal{I}$, is a constant vector in $\m{E}_{\rm R}$, which we denote as $\lsc{\u{E}}_i$.
The position vectors $\pr{\lsc{\u{X}}}_{\ell\in\mathcal{J}}$ and the
directions $\pr{\lsc{\u{E}}_{i}}_{\ell\in\mathcal{J}, i\in\mathcal{I}}$ are known from the arrangement and orientation of the accelerometers
at the experiment's beginning.

For $\ell\in\mathcal{J}$, let $\lsc{\busf{A}}(\tau):=\pr{\lsc{\alpha}_{j}(\tau)\lsc{\u{E}}_{j}\cdot\u{E}_{i}}_{i\in\mathcal{I}}$ (no sum over $\ell$), where $\lsc{\alpha}_{i}(\tau)$, $i\in\mathcal{I}$, is the measurement reported by accelerometer $\lsc{\u{X}}$ for the (non-dimensional) component of its acceleration in the $\lsc{\u{e}}_{\tau i}$ direction\footnote{\label{fn:Or-to-be}Or to be mathematically precise, in the $\lsc{\u{a}}_{\tau i}\in\m{A}$
direction that is defined such that $\pr{\lsc{\u{a}}_{\tau i}\u{s}}\u{s}=\lsc{\u{e}}_{\tau i}$.} at the time instance $\tau$.
And let $\lsc{\usf{X}}:=\pr{\lsc{\u{X}}\cdot\u{E}_i}_{i\in\mathcal{I}}$.
Then, we compute $\usf{P}(\tau)^{\rm p}$ as $\usf{P}(\tau)$ is
estimated in \cite{rahaman2019} using the equation
\begin{equation}
\usf{P}(\tau)^{{\rm p}}=\pr{\tps{\lsc[i]{\Delta\busf{A}}(\tau)}\,\pr{\lsc[j]{\dsf{X}}}}\,\pr{\pr{\lsc[i]{\dsf{Y}}}\,\tps{\lsc[j]{\dsf{Y}}}},\,\label{equ:p}
\end{equation}
where $\lsc[i]{\Delta\busf{A}}(\tau):=\lsc[i+1]{\busf{A}}(\tau)-\lsc[1]{\busf{A}}(\tau)$, $\lsc[i]{\Delta\usf{X}}:=\lsc[i+1]{\usf{X}}-\lsc[1]{\usf{X}}$.
The ordered sets $\lsc[i]\dsf{Y}$ belong to $\mathcal{M}_{\nsd,1}(\mathbb{R})$
and are defined by the equation

\begin{equation}
\pr{\lsc[1]{\Delta\usf{Y}},\ldots,\lsc[\nsd]{\Delta\usf{Y}}}=\tipsb{\pr{\lsc[1]{\Delta\usf{X}},\ldots,\lsc[\nsd]{\Delta\usf{X}}}},
\end{equation}
where $\tipsb{\pr{\cdot}}$ is the operator that acts on an invertable
element of $\mathcal{M}_{\nsd,\nsd}(\mathbb{R})$ and returns the transpose
of its inverse.

\paragraph{Corrector step\label{par:Corrector-step}}

In \ref{subsec:A-spectral-decomposition} we show that $\text{sym}(\usf{P}(\tau))$ allows itself to be decomposed as
\begin{subequations}
\begin{align}
&\usf{N}(\tau)\,\usf{D}(\tau)\,\tpsb{\usf{N}}(\tau),\label{equ:SymPTheoreticalSpecDecomposition_P}\intertext{where $\usf{N}(\tau)\in\mathcal{M}_{\nsd,\nsd}(\mathbb{R})$ is an orthogonal matrix, i.e.,} &\tpsb{\usf{N}}(\tau)\usf{N}(\tau)=\usf{I},\label{equ:SymPTheoreticalSpecDecomposition_N}
\intertext{and $\usf{D}(\tau)\in\mathcal{M}_{\nsd,\nsd}(\m{R})$ is a diagonal matrix that for $\nsd=2$ and $3$, respectively, has the form} &\usf{D}(\tau)=\text{diag}\pr{-\lambda(\tau)^2,-\lambda(\tau)^2}~\text{and} ~\text{diag}\pr{0,-\lambda(\tau)^2,-\lambda(\tau)^2},
\label{equ:SymPTheoreticalSpecDecomposition_D}
\end{align}
\label{equ:SymPTheoreticalSpecDecomposition}
\end{subequations}
where $\lambda(\tau)\in\m{R}$ and the function $\text{diag}(\cdot):\m{F}^{\nsd}\to\c{M}_{\nsd,\nsd}(\m{F})$,
where $\m{F}$ is either $\m{R}$ or $\m{C}$, is defined such that $\text{diag}(a_1,\ldots, a_{\nsd})$ is a diagonal matrix with diagonal entries $a_1,\ldots,a_{\nsd}$.

The matrix $\text{sym}(\usf{P}(\tau))$ allowing the decomposition $\eqref{equ:SymPTheoreticalSpecDecomposition}$ is critical for carrying out \ref{enu:sqrtAO-Step2} of the $\sqrt{\text{AO}}$-algorithm.
In an ideal scenario, in which there are no experimental errors or
noise in the accelerometer measurements, $\usf{P}(\tau)^{\rm p}$ would
be the same as $\usf{P}(\tau)$.
However, due to the experimental noise and other errors $\usf{P}(\tau)^{\rm p}$
will generally be different from $\usf{P}(\tau)$.
In general, such a deviation would not be of much consequence, since,
experimental measurements of physical quantities, more often than
not, are different from the true values of those quantities.
Thus, generally, we would, as done by \cite{rahaman2019}, take $\usf{P}(\tau)^{\rm p}$ to be the final estimate for $\usf{P}(\tau)$ and no longer distinguish between $\usf{P}(\tau)^{\rm p}$ and $\usf{P}(\tau)$.
However, in the present case the deviation of $\usf{P}(\tau)^{\rm p}$
from $\usf{P}(\tau)$ has an important consequence which requires us
to not take $\usf{P}(\tau)^{\rm p}$ as $\usf{P}(\tau)$'s final estimate.
The important consequence is that in general $\text{sym}(\usf{P}(\tau)^{\rm p})$ will not allow a decomposition of the form \eqref{equ:SymPTheoreticalSpecDecomposition}.
In general, it will only allow itself to be decomposed as $\usf{N}_{\mathscr{p}}(\tau)\,\text{diag}\pr{\lambda_i(\tau),\ldots,\lambda_{\nsd}(\tau)}\,\tpsb{\usf{N}_{\mathscr{p}}}(\tau)$, where $\usf{N}_{\mathscr{p}}(\tau)$'s columns are the eigenvectors
of $\text{sym}(\usf{P}(\tau)^{\rm p})$ that are chosen such that $\usf{N}_{\mathscr{p}}(\tau)$ is orthogonal and their corresponding eigenvalues $\lambda_{i}(\tau)\in\m{R}$ form a non-increasing sequence, i.e., $\lambda_{1}(\tau)\ge\ldots\ge\lambda_{\nsd}(\tau)$.
Therefore, instead of taking $\usf{P}(\tau)^{\rm p}$ as the final
estimate of $\usf{P}(\tau)$ we derive the final estimate for $\usf{P}(\tau)$,
as we detail next, using $\usf{P}(\tau)^{\rm p}$ so that its symmetric
part does allow a decomposition of the form $\eqref{equ:SymPTheoreticalSpecDecomposition}$.

We take the skew-symmetric part of our final estimate for $\usf{P}(\tau)$
to be the same as that of $\usf{P}(\tau)^{\rm p}$.
We take its symmetric part to be

\begin{subequations}
\begin{align}
&\usf{N}_{\mathscr{p}}(\tau)\,\cusf{D}(\tau)\,\tpsb{\usf{N}}_{\mathscr{p}}(\tau),
\intertext{where}
\cusf{D}(\tau)&:=\text{diag}\pr{0,-\check{\lambda}(\tau)^2,-\check{\lambda}(\tau)^2},
\label{equ:DConstruction_D_nsd3}
\intertext{with}
\check{\lambda}(\tau)&:=\begin{cases}\sqrt{{-\frac{\lambda_{2}(\tau)+\lambda_{3}(\tau)}{2}}}, &\lambda_{2}(\tau)+\lambda_{3}(\tau)\leq0,\\0, &\lambda_{2}(\tau)+\lambda_{3}(\tau)>0,\end{cases}
\label{equ:DConstruction_lambda_nsd3}
\intertext{for $\nsd=3$, and}
\cusf{D}(\tau)&:=\text{diag}\pr{-\check{\lambda}(\tau)^2,-\check{\lambda}(\tau)^2},\label{equ:DConstruction_D_nsd2}
\intertext{with}
\check{\lambda}(\tau)&:=
\begin{cases}
\sqrt{{-\frac{\lambda_{1}(\tau)+\lambda_{2}(\tau)}{2}}}, &\lambda_{1}(\tau)+\lambda_{2}(\tau)\leq0,\\0, &\lambda_{1}(\tau)+\lambda_{2}(\tau)>0,
\end{cases}
\label{equ:DConstruction_lambda_nsd2}
\intertext{for $\nsd=2$.}\notag
\end{align}
\label{equ:FinalEstimateSymPDecomposition}
\end{subequations}

The orthogonal matrix $\usf{N}_{\mathscr{p}}(\tau)$ and the eigenvalues $\lambda_{i}(\tau)$ can be obtained from the spectral or symmetric-Schur \cite[\S{8}]{golub2013} decomposition of $\text{sym}(\usf{P}(\tau)^{\rm p})$.
Since $\text{sym}(\usf{P}(\tau)^{\rm p})$ is a real symmetric matrix, it is always possible to carry out $\text{sym}(\usf{P}(\tau)^{\rm p})$'s spectral or symmetric Schur decomposition.

To summarize, we take $\usf{P}(\tau)$'s final estimate to be

\begin{subequations}
\begin{align}
&\usf{P}^{\rm p}(\tau)+\Delta\usf{P}(\tau),
\intertext{where}
\Delta\usf{P}(\tau)&:=\text{sym}\pr{\cusf{P}(\tau)-\usf{P}^{\rm p}(\tau)},\\
\cusf{P}(\tau)&:=\usf{N}_{\mathscr{p}}(\tau)\check{\usf{D}}(\tau)\tpsb{\usf{N}}_{\mathscr{p}}(\tau).
\end{align}
\label{equ:PConstruction}
\end{subequations}

It can be ascertained that the symmetric
part of our final estimate for $\usf{P}(\tau)$, namely $\cusf{P}(\tau)$,
allows a decomposition of the form $\eqref{equ:SymPTheoreticalSpecDecomposition}$.
In fact, that decomposition is precisely the one given by \eqref{equ:FinalEstimateSymPDecomposition}.
For $\nsd=2$ or $3$ it can be shown that $\cusf{P}(\tau)$ is the best
approximation\footnote{We plan to publish this result along with its proof elsewhere.}, in the Frobenius norm, to $\text{sym}\pr{\usf{P}^{\rm p}(\tau)}$ in the set of $\nsd\times\nsd$
real symmetric negative-semidefinite matrices whose negative eigenvalues (when they exist) are of even algebraic multiplicities.

\subsubsection{Estimating $\usf{q}(\tau)$\label{subsec:Estimating-1}}

After estimating $\usf{P}(\tau)$
as described in $\S$\ref{subsec:Estimating} using $\eqref{equ:PConstruction}$
we estimate $\usf{q}(\tau)$ as
\begin{equation}
\lsc{\bar{\usf{A}}}(\tau)-\boldsymbol{\sf P}(\tau)\,\lsc{\usf{X}},
\end{equation}
where $\mathscr{l}$ is some particular integer in $\mathcal{J}$.

\subsection{$\sqrt{\text{AO}}$-algorithm, \ref{enu:sqrtAO-Step2} of $3$}

\label{step2}

In \ref{subsec:Calculation-of-} we show using $\usf{P}(\tau)$'s
decomposition $\eqref{equ:SymPTheoreticalSpecDecomposition}$ that $\busf{W}(\tau)$ can be computed from $\eqref{equ:sym-1}$ as

\begin{equation}
\busf{W}(\tau)=
\pm
\begin{cases}
\usf{N}(\tau)
\star\pr{\lambda(\tau)}
\tpsb{\usf{N}}(\tau),&\nsd=2,
\\
\usf{N}(\tau)
\,\star\pr{\pr{\lambda(\tau),0,0}}\,
\tpsb{\usf{N}}(\tau),&\nsd=3,
\end{cases}
\label{equ:WbTheoreticalDecomposition}
\end{equation}
where the map $\star\pr{\cdot}$ is defined in \ref{subsec:hodgestar}.

Using the decomposition $\eqref{equ:FinalEstimateSymPDecomposition}$ for the symmetric part of our final estimate for $\usf{P}(\tau)$ and similar calculations as those used in $\S$\ref{subsec:Calculation-of-}, it can be shown that if we compute our final estimate for $\busf{W}(\tau)$ as

\begin{equation}
\pm
\begin{cases}
\usf{N}_{\mathscr{p}}(\tau)
\star\pr{\check{\lambda}(\tau)}
\tpsb{\usf{N}}_{\mathscr{p}}(\tau),
&\nsd=2,
\\
\usf{N}_{\mathscr{p}}(\tau)
\,\star\pr{\pr{\check{\lambda}(\tau),0,0}}\,
\tpsb{\usf{N}}_{\mathscr{p}}(\tau), &\nsd=3,
\end{cases}
\label{equ:WbEstimateDecomposition}
\end{equation}
then it and our final estimate for $\usf{P}(\tau)$ will satisfy $\eqref{equ:sym-1}$.

We take the time discrete versions of the $\busf{W}$ and $\usf{P}$ maps to be constant over each time interval $\Delta\tau_{n}:=[n\Delta\tau,(n+1)\Delta\tau)$, where $n\in (0,1,\cdots)$ and $\Delta\tau\in\mathbb{R}$.
We denote the values of these two maps over $\Delta\tau_{n}$ as $\busf{W}(n)$
and $\usf{P}(n)$, respectively.
The quantity $\busf{W}(0)$ is known from initial conditions.
For $n>0$ we compute $\busf{W}(n)$ using \eqref{equ:WbEstimateDecomposition}.
Using the positive and negative signs in \eqref{equ:WbEstimateDecomposition} will give us two different estimates for $\busf{W}(n)$.
Among those two estimates we choose the one that is closer to $\busf{W}$'s value over the previous time interval.
To be precise, we choose the estimate that gives a lower value for
the metric $m(\busf{W}(n),\busf{W}(n-1))$, where $m:\mathfrak{so}(\mathbb{R},\nsd)^2\to\m{R}$,
\begin{equation}
m(\busf{W}(n),\busf{W}(n-1))=\arccos\left(\frac{\star\pr{\busf{W}(n)}\cdot\star\pr{\busf{W}(n-1)}}{\lVert\star\pr{\busf{W}(n)}\rVert\lVert\star\pr{\busf{W}(n-1)}\lVert}\right).
\label{equ:realW}
\end{equation}

The metric $\eqref{equ:realW}$ is a measure of the difference between
between $\busf{W}(n)$ and $\busf{W}(n-1)$.
Our criterion for choosing between the two estimates given by $\eqref{equ:WbEstimateDecomposition}$
is essentially based on the assumption that in most practical scenarios $\busf{W}$ should be a continuous function
of time, and on the ansatz that due to the continuity of $\busf{W}$
the true $\busf{W}(n)$ would be the one that is closer to $\busf{W}(n-1)$
when $\norm{\star\pr{\busf{W}(n-1)}}$ is large.

If $\busf{W}(n-1)=\u{0}$ or when $\norm{\star\pr{\busf{W}(n-1)}}$ is small, then our above criterion for choosing between the two estimates for $\busf{W}(n)$ cannot be used.
In such cases we first derive a prediction for $\busf{W}(n)$ by applying the AO-algorithm to the previous time interval and then choose the estimate that is
closer to that prediction.

\subsection{$\sqrt{\text{AO}}$-algorithm, \ref{enu:sqrtAO-Step3} of $3$}
\label{step3}

We use a slightly modified version of the numerical integration scheme
described by equations $3.15$, $3.16$, and $3.17$ in \cite[\S3.2]{rahaman2019} to solve $\eqref{equ:rotation}$.
As we did with $\busf{W}$ and $\usf{P}$, we assume that the discrete version
of $\usf{Q}$ remains constant over each time interval $\Delta\tau_n$
and denote its constant values as $\usf{Q}(n)$, where $n\in (0,1,\cdots)$.
The matrix $\usf{Q}(0)$ is taken to be known from the initial conditions
of the experiment.
For $n>0$ the matrix $\usf{Q}(n)$ is computed as

\begin{subequations}
\begin{align}
\usf{Q}(n)&=\usf{Q}(n-1)\,\mathsf{e}^{\Delta\tau\,\overline{\boldsymbol{\sf W}}_{n-\frac{1}{2}}},
\label{equ:NISchemeForSolRotation_QUpdate}
\intertext{where the map $\mathsf{e}^{\pr{\cdot}}:\mathfrak{so}(\mathbb{R},\nsd)\to SO(\nsd)$ is defined by the equation\footnotemark}
\mathsf{e}^{\pr{\cdot}}&=\usf{I}+\text{sinc}\pr{\left\lVert\star\pr{\cdot}\right\rVert }\pr{\cdot}+\frac{1}{2}\pr{\text{sinc}\pr{\frac{\left\lVert {\star\pr{\cdot}}\right\rVert }{2}}}^{2}\pr{\cdot}^{2},
\label{equ:NISchemeForSolRotation_MatExpDef}
\intertext{and}
\busf{W}_{n-\frac{1}{2}}&:=\frac{1}{2}\pr{\busf{W}(n)+\busf{W}(n-1)}.
\label{equ:NISchemeForSolRotation_WbUpdate}
\end{align}
\label{equ:NISchemeForSolRotation}
\end{subequations}
\footnotetext{This equation is the corrected version of equation $3.17$ in \cite[\S{3.2}]{rahaman2019}, which has two typos in it.}

The difference between the integration scheme $\eqref{equ:NISchemeForSolRotation}$ and that given by $3.15$, $3.16$,
and $3.17$ in \cite[\S3.2]{rahaman2019} is the manner in which $\busf{W}_{n-\frac{1}{2}}$ is computed.
In Rahaman \textit{et al.}'s integration scheme, $\busf{W}_{n-\frac{1}{2}}$ is computed as $\busf{W}(n-1)+\frac{\Delta\tau}{2}\text{skew}\pr{\usf{P}(n-1)}$, whereas we compute it using $\eqref{equ:NISchemeForSolRotation_WbUpdate}$.
Here we use $\text{skew}\pr{\usf{P}(n-1)}$ to denote the skew-symmetric part of the matrix $\usf{P}(n-1)$.

\section{\textit{In silico} validation, evaluation and comparison of the $\sqrt{\text{AO}}$-algorithm}
\label{sec:In-silico-validation}

In this section, we check the validity and robustness of the $\sqrt{\text{AO}}$-algorithm.
We do that by feeding in virtual accelerometer data, to which differing amounts
of bias and noise type errors have been added, to the $\sqrt{\text{AO}}$
and $\text{AO}$ algorithms, and comparing their resulting predictions.
We discuss the creation of the virtual accelerometer data in $\S$\ref{subsec:Rigid-ellipsoid-impact}, and simulating bias and noise type errors in $\S$\ref{subsec:Simulating-noise-and}.
We compare the predictions in $\S$\ref{subsec:Comparison-of-AO-alogrithm}.

\subsection{Virtual accelerometer data from the simulation of a rigid ellipsoid
impacting an elastic half-space}
\label{subsec:Rigid-ellipsoid-impact}

The virtual accelerometer data we use for the comparison is from the
numerical simulation of a rigid ellipsoid impacting an elastic half-space.
This simulation is presented and discussed in detail in \cite{rahaman2019}, starting in $\S 4$.
However, for the readers convenience we give a very brief description
of that simulation here.

\begin{figure}[ht]
\centering{}
\includegraphics[width=0.85\textwidth]{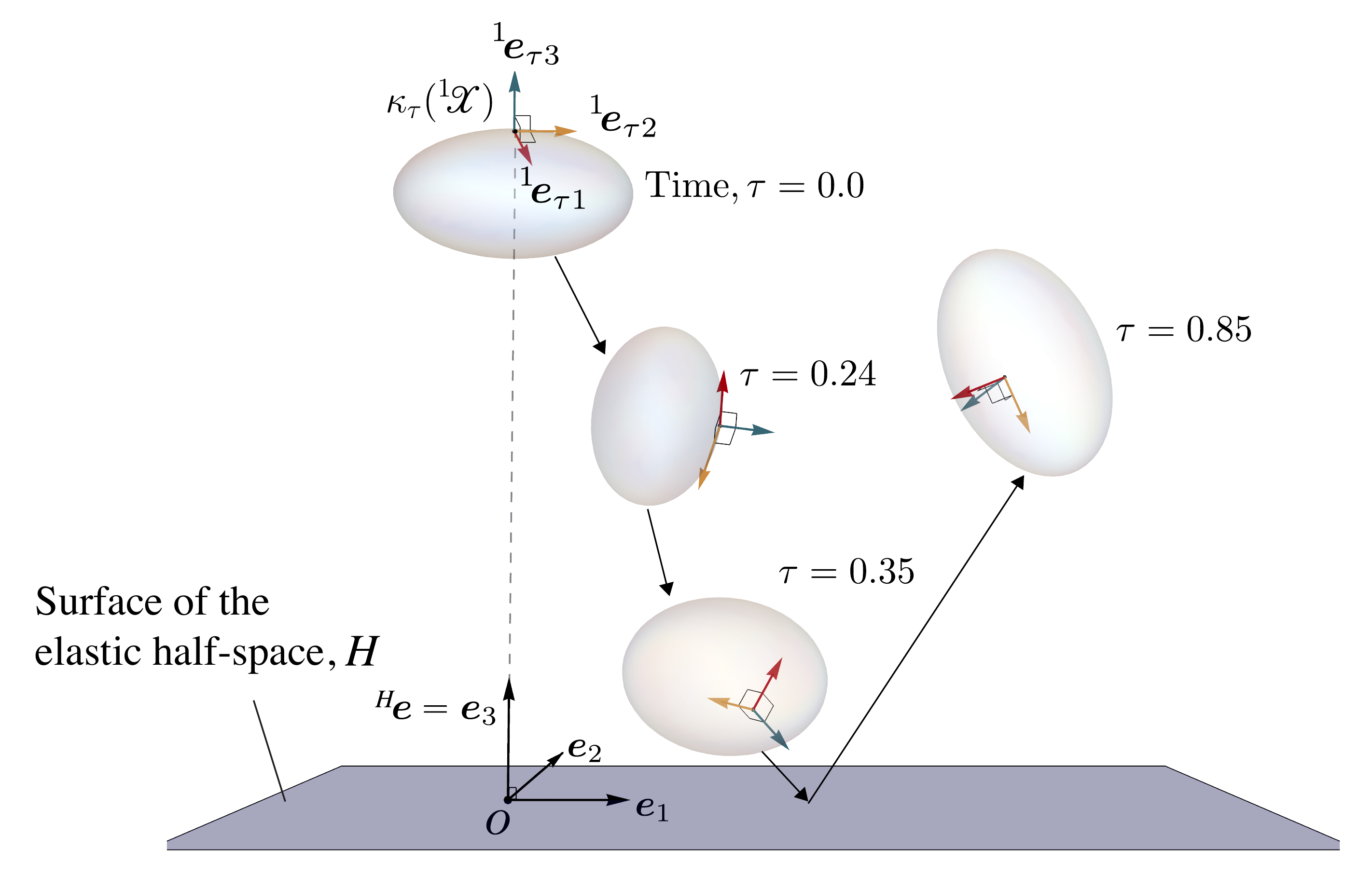}
\caption{Configuration of the rigid ellipsoid at different time instances in the simulation
of it impacting an elastic half space  (see \S\ref{subsec:Rigid-ellipsoid-impact} for details).
In the simulation, the ellipsoid, $\mathcal{B}$, is dropped onto an elastic half-space, $H$,
under the action of gravity with the initial angular and translational velocities prescribed.
The ellipsoid's initial position in $\mathcal{E}$ is prescribed by taking $\usf{c}(0)=\pr{0,0,0.75}$, and $\usf{Q}(0)=\text{diag}\pr{1,1,1}$.
Its initial velocities are prescribed by setting $\star(\usf{W}(0))=\pr{5,5,5}$, and $\usf{c}'(0)=\pr{0.75,0,0}$. (modified from \cite{rahaman2019}, copyright 2020, Elsevier)}
\label{fig:impact}
\end{figure}

In the simulation an ellipsoid, $\mathcal{B}$, is dropped onto an elastic half-space, $H$, under the action of gravity with the initial angular and translational velocities prescribed (see Fig.~\ref{fig:impact}).
In the simulation the Euclidean point space $\mathcal{E}$, in which
the ellipsoid and the half-space, respectively, execute their motion
and deformation, is taken to be three dimensional, i.e., $\nsd=3$.
The vectors $\u{E}_i$ and $\u{e}_i$, $i\in\mathcal{I}$, are taken to have units of meters and $\u{s}$ to have units of seconds.
Hence $\u{v}_i$ and $\u{a}_i$, $i\in\mathcal{I}$, have units of meters-per-second and meters-per-second-squared, respectively.

\begin{figure}[ht]
\centering{}
\includegraphics[width=0.85\textwidth]{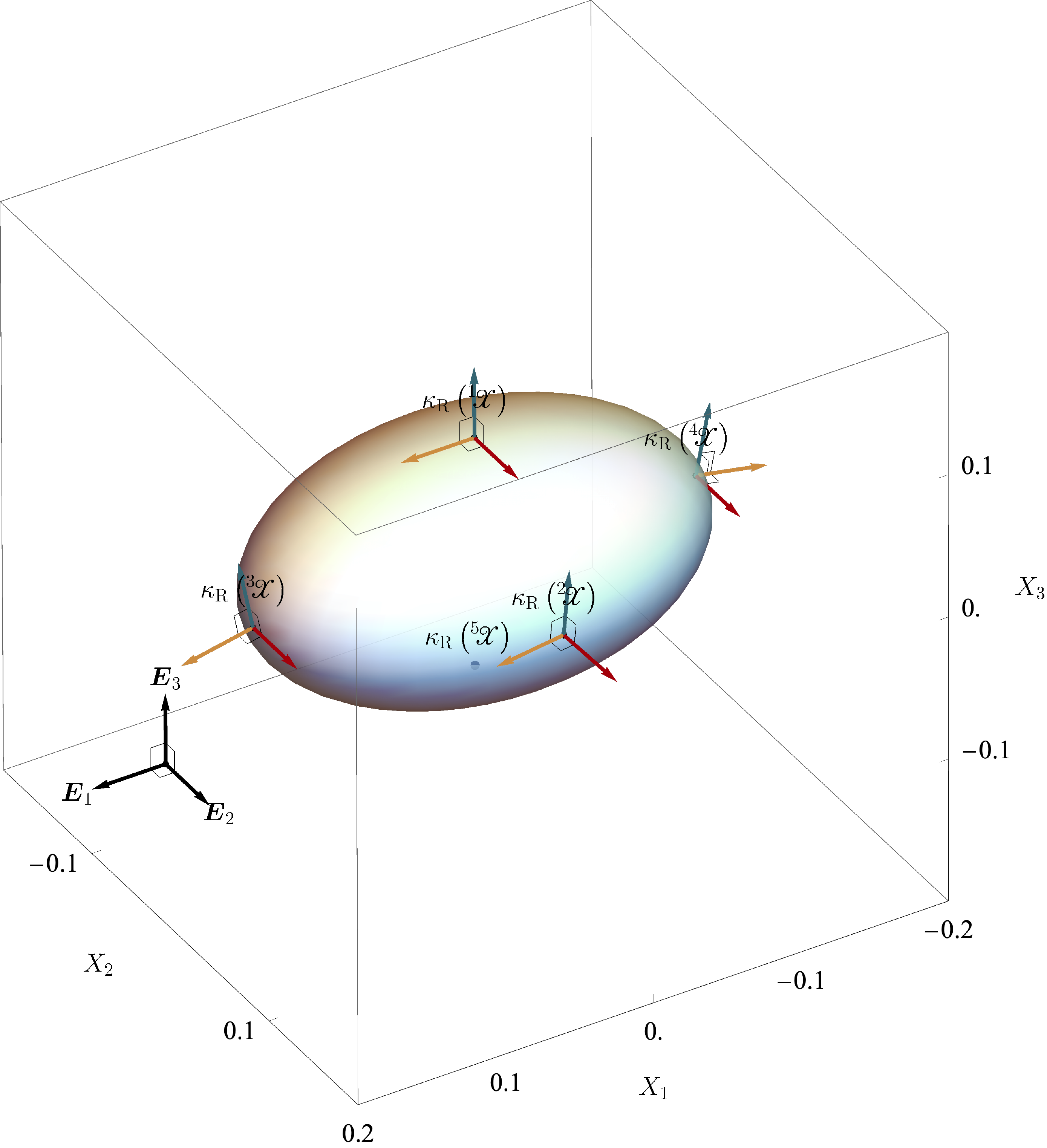}
\caption{Accelerometer arrangement and orientation in numerical simulation of a rigid ellipsoid impacting an elastic half-space (see \S\ref{subsec:Rigid-ellipsoid-impact} for details).
In the reference point space $\mathcal{E}_{\rm R}$ the ellipsoid, $\mathcal{B}$, occupies
the region $\set{(X_1,X_2,X_3)}{\pr{X_1/a}^2+\pr{X_2/b}^2+\pr{X_3/c}^2\le 1}$,
where $(a,b,c)=(0.15,0.10,0.08)$.
Four virtual accelerometers are, respectively, attached to the ellipsoid's material particles $\lsc{\c{X}}$, $\ell\in\mathcal{J}$.
The reference position vectors of $\lsc{\u{X}}$, $\ell\in\mathcal{J}$,
are, respectively, $c\u{E}_3$, $b\u{E}_2$, $a\u{E}_1$, and $-a\u{E}_1$,
where $\pr{\u{E}_i}_{i\in\mathcal{I}}$ are shown in the figure as well.
The accelerometers' orientations are given by $\pr{\lsc{\u{E}}_i}_{i\in\mathcal{I}}$, $\ell\in\mathcal{J}$.
The component representation of  $\pr{\lsc[1]{\u{E}}_i}_{i\in\mathcal{I}}$ w.r.t. $\pr{\u{E}_i}_{i\in\mathcal{I}}$ is $\pr{\pr{0,1,0},\pr{1,0,0},\pr{0,0,1}}$; of $\pr{\lsc[2]{\u{E}}_i}_{i\in\mathcal{I}}$ is
$\pr{\pr{-\frac{2}{\sqrt{229}},\frac{225}{229},-\frac{30}{229}},\pr{\frac{15}{\sqrt{229}},\frac{30}{229},-\frac{4}{229}},\pr{0,\frac{2}{\sqrt{229}},\frac{15}{\sqrt{229}}}}$;
of $\pr{\lsc[3]{\u{E}}_i}_{i\in\mathcal{I}}$ is $\pr{\pr{0,1,0},\pr{\frac{5}{\sqrt{26}},0,-\frac{1}{\sqrt{26}}},\pr{\frac{1}{\sqrt{26}},0,\frac{5}{\sqrt{26}}}}$; and of
$\pr{\lsc[4]{\u{E}}_i}_{i\in\mathcal{I}}$ is $\pr{\pr{0,1,0},\pr{-\frac{5}{\sqrt{26}},0,-\frac{1}{\sqrt{26}}},\pr{-\frac{1}{\sqrt{26}},0,\frac{5}{\sqrt{26}}}}$.
We apply the $\sqrt{\text{AO}}$-algorithm to the accelerometer
data from the four virtual accelerometers $\lsc{\c{X}}$, $\ell\in\mathcal{J}$ to predict the acceleration of the material particle $\lsc[5]{\c{X}}$.
The reference position vector of $\lsc[5]{\c{X}}$ is $-c\u{E}_3$. (modified from \cite{rahaman2019}, copyright 2020, Elsevier)}
\label{fig:accposition}
\end{figure}

The reference configuration of the ellipsoid is given in Fig.~\ref{fig:accposition}.
In $\mathcal{E}_{\rm R}$ the ellipsoid occupies the region $\set{(X_1,X_2,X_3)}{\pr{X_1/a}^2+\pr{X_2/b}^2+\pr{X_3/c}^2\le 1}$,
where $(a,b,c)=(0.15,0.10,0.08)$.
The half-space when it is undeformed in $\mathcal{E}$ occupies the
region $x_3< 0$.
The initial location and orientation of $\mathcal{B}$ w.r.t. $H$ in $\mathcal{E}$ are shown in Fig.~\ref{fig:impact}.
They correspond to the initial conditions $\usf{Q}(0)=\text{diag}\pr{1,1,1}$ and $\usf{c}(0)=\pr{0,0,0.75}$.

The mechanics of $H$ is modeled using the theory of small deformation
linear elasto-statics and taking $H$'s Young's modulus and Poisson's
ratio to be $10^4~\rm Pa$ and $0.3$, respectively.
The ellipsoid is rigid and homogeneous.
Its density and total mass are $1989.44~\rm Kg/m^3$ and $10~\rm Kg$,
respectively.

The ellipsoid's dynamics are obtained by numerically solving its linear
and angular momentum balance equations.
The force in those equations arises due to the action of gravity on $\mathcal{B}$ and $\mathcal{B}$'s interaction with $H$; and the torque exclusively from $\mathcal{B}$'s interaction with $H$.
The interaction between $\mathcal{B}$ and $H$ is modeled using the
Hertz contact theory, e.g., \cite{Kesari2011,Kesari2011-2}.
For details on effecting a numerical solution to the balance equations
see \cite[\S B1.1]{rahaman2019}.
For more details regarding the contact modeling see \cite[\S B1.2]{rahaman2019}.

\begin{figure}[ht]
\centering{}
\includegraphics[width=0.8\textwidth]{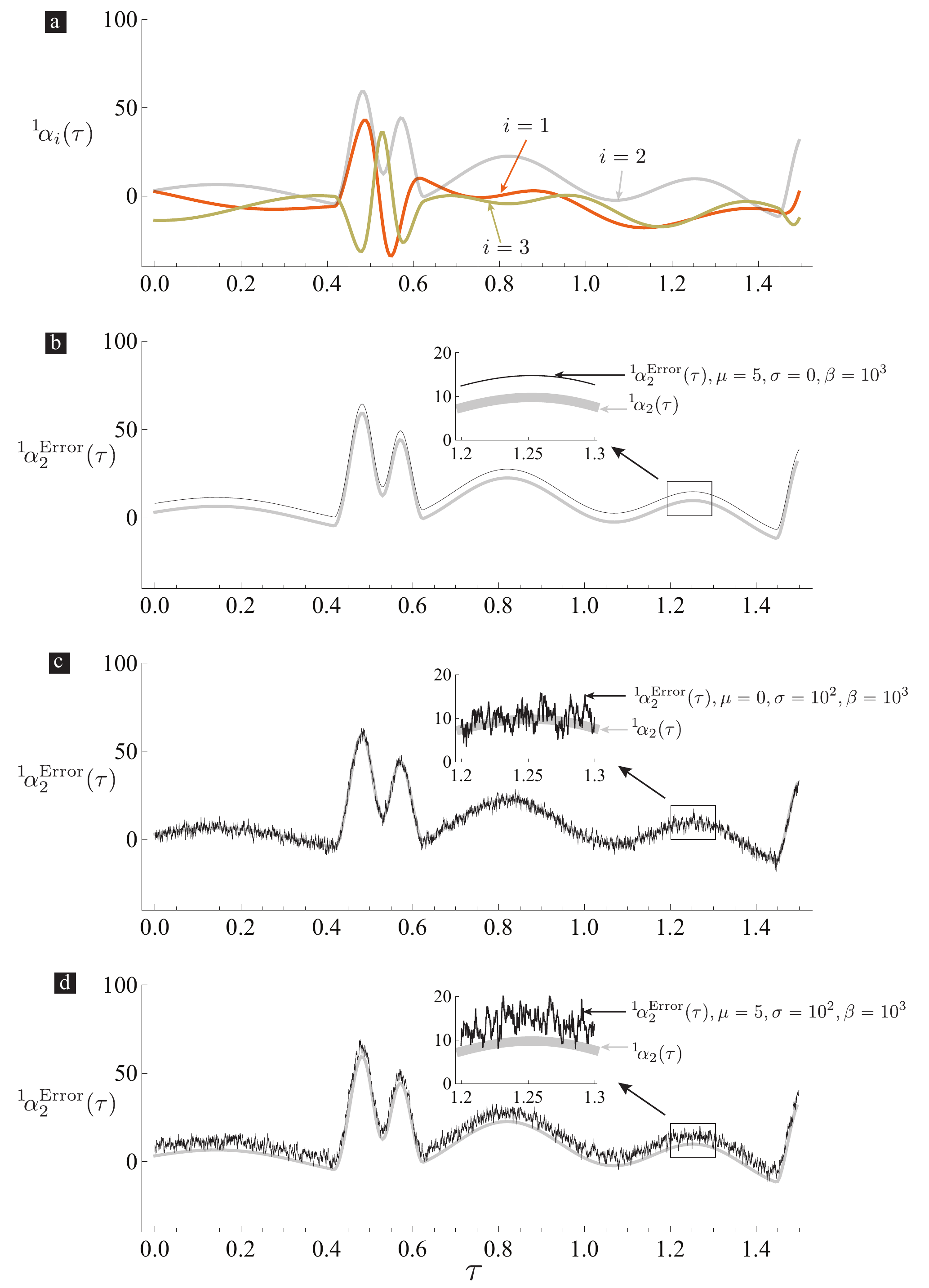}
\caption{The acceleration components $\lsc[1]{\alpha}_{i}(\tau)$, $i\in\mathcal{I}$, of the virtual accelerometer $\lsc[1]\c{X}$ before and after addition
of synthetic errors (see \S\ref{subsec:Simulating-noise-and} for details). 
(a) shows the acceleration components before the addition of synthetic errors. 
(b)--(d) show the error-inclusive acceleration component  $\lsc[1]\alpha_{2}^{\rm Error}$, which is generated by adding different errors to the acceleration component $\lsc[1]\alpha_{2}$. 
In (b), (c), and (d) the error time signals are particular realizations of the OU process for the OU parameter sets $(\mu, \sigma, \beta)=(5,0,10^3)$, $(0,10^2,10^3)$, and $(5,10^2,10^3)$, respectively. 
The error in  (b) corresponds to \ref{category1} (exclusively bias type errors); in (c) to  \ref{category2} (exclusively noise type errors); and in (d) to \ref{category3} (a combination of bias and noise type errors). 
}
\label{fig:accinput}
\end{figure}

Four virtual accelerometers are taken to be rigidly attached to the
ellipsoid's material particles $\lsc{\c{X}}$, $\ell\in\mathcal{J}$.
The locations and orientations of those accelerometers w.r.t. $\mathcal{B}$
in $\mathcal{E}_{\rm R}$ are shown in Fig.~\ref{fig:accposition}.
Their position vectors $\lsc{\u{X}}$ and orientations $\pr{\lsc{\u{E}}_i}_{i\in\mathcal{I}}$, $\ell\in\mathcal{J}$, are given in the caption of Fig.~\ref{fig:accposition}.
The acceleration of any of the ellipsoid's material particles can
be obtained from the simulation results using the procedure outlined
in \cite[\S B2]{rahaman2019}.
For $i\in\mathcal{I}$, the values of $\lsc[1]{\alpha}_i$, which is
the component of $\lsc[1]{\c{X}}$'s acceleration in the $\lsc[1]{\u{e}_{\tau i}}$ direction, or to be more precise the $\lsc[1]{\u{a}_{\tau i}}$ direction\footnoteref{fn:Or-to-be} (see Fig.~\ref{fig:impact}), at a sequence of time instances are shown in Fig.~\ref{fig:accinput}(a).

\subsection{Adding synthetic errors to virtual accelerometer data\label{subsec:Simulating-noise-and}}

The acceleration components $\lsc{\alpha}_{i}$, $\ell\in\mathcal{J}$, from the simulation do not contain any errors; other than, of course,
the errors that arise due to numerical discretization of the balance
equations, numerical round-off, etc. However,
those type of errors are of insignificant magnitude. Using
the error free virtual accelerometer
data $\lsc{\alpha}_{i}$, $\ell\in\mathcal{J}$, from the simulation
we generate virtual error-inclusive accelerometer data $\lsc{\alpha}_{i}^{\rm Error}$, $\ell\in\mathcal{J}$, as

\begin{equation}
\lsc{\alpha_{i}^{\rm Error}(\tau)}=\lsc{\alpha_{i}(\tau)}+\eta_{\tau}.
\label{eq:error}
\end{equation}

In equation $\eqref{eq:error}$ $\eta_{\tau}$
denotes a particular realization of the Ornstein-Uhlenbeck (OU) process \cite{dimian2014}.
We will describe shortly what we mean by a ``realization''.
The OU process is a continuous time and state stochastic process that
is defined by the integral equation
\begin{equation}
\eta_{\tau_{1}+\tau_{2}}-\eta_{\tau_{1}}=\beta\int_{\tau_{1}}^{\tau_{1}+\tau_{2}}\left(\mu-\eta_{\tau}\right)\,d\tau+\sigma\int_{\tau_{1}}^{\tau_{1}+\tau_{2}}\,dW_{\tau},
\label{eq:OUprocess}
\end{equation}
where the second integral on the right is an It\^{o} integral and $W_{\tau}$
is the Wiener process \cite{Bibbona2008}.
The real number $\mu$ is called the mean value, $\sigma\geq0$ the
diffusion coefficient, and $\beta>0$ the drift coefficient. The
symbols $\tau_{1}$, $\tau_{2}$ denote any two (non-dimensional) time
instances.
Since the OU process is a stochastic process, a given set of OU parameters, i.e., a particular set of $\mu$, $\sigma$, and $\beta$ values, define an entire
family or population of real valued functions on $\mathbb{R}$.
For a given OU parameter set, a particular realization of the OU process
is obtained by drawing $\eta_0$ from a Gaussian distribution of mean $\mu$
and variance $\sigma^2/\pr{2\beta}$ and solving $\eqref{eq:OUprocess}$.
As a consequence of $\eqref{eq:error}$, $\lsc{\alpha}_{i}^{\rm Error}$, $\ell\in\mathcal{J}$, too are stochastic processes.

For $i\in\mathcal{I}$ and $\ell\in\mathcal{J}$, when $\mu\neq0$
and $\sigma = 0$, any particular realization of $\lsc{\alpha}_{i}^{\rm Error}$
will contain only bias type errors.
A representative realization of $\lsc[1]{\alpha}_{2}^{\rm Error}$ for $\mu=5$, $\sigma=0$, and $\beta=10^3$ is shown in Fig.~\ref{fig:accinput}(b). Alternatively, when $\mu=0$ and $\sigma\neq 0$ any particular realization of $\lsc{\alpha}_{i}^{\rm Error}$ will only contain noise type errors.
A representative realization of $\lsc[1]{\alpha}_{2}^{\rm Error}$
for $\mu=0$, $\sigma=10^2$, and $\beta=10^3$ is shown in Fig.~\ref{fig:accinput}(c). In
general when $\mu$ and $\sigma$ are both non-zero, realizations of $\lsc{\alpha}_{i}^{\rm Error}$ will contain both bias and noise type errors.
A representative realization of $\lsc[1]{\alpha}_{2}^{\rm Error}$ for $\mu=5$, $\sigma=10^2$, and $\beta=10^3$ is shown in Fig.~\ref{fig:accinput}(d).

From here on unless otherwise specified the value of $\beta$ will
always be equal to $10^3$.

\subsection{Comparison of $\sqrt{\text{AO}}$ and AO algorithms using virtual
error-inclusive accelerometer data\label{subsec:Comparison-of-AO-alogrithm}}

We compare the predictions of the $\sqrt{\text{AO}}$ and AO algorithms
for the following categories of OU parameter sets.

\begin{enumerate}[label=\textit{Category~\Roman*},leftmargin=2cm]
\item Exclusively bias type errors:  $\mu=0$, $0.1$, $0.2$, $0.5$, and $1$, and $\sigma=0$ (see Table.~\ref{Tab:mu_sigma0}).\label{category1}
\item Exclusively noise type errors:  $\mu=0$,  and $\sigma=0$, $1$, $10$, $50$, and $10^2$ (see Table.~\ref{Tab:sigma}).\label{category2}
\item Both bias and noise type errors: $\mu=0$, $0.1$, $0.2$, $0.5$, and $1$, and $\sigma=10$ (see Table.~\ref{Tab:mu}).\label{category3}
\end{enumerate}

For a given OU parameter set we generate a large number of $\lsc{\alpha}_{i}^{\rm Error}$, $\ell\in\mathcal{J}$, realizations.
We apply the $\sqrt{\text{AO}}$ and $\text{AO}$ algorithms to each of those
realizations and derive a population of predictions for the acceleration
of the material particle $\lsc[5]{\c{X}}$ (see Fig.~\ref{fig:accposition}).
We denote the error-free (non-dimensional) acceleration of $\lsc[5]{\mathcal{X}}$, which we know from the rigid-ellipsoid-impact-simulation's results, at the time instance $\tau$ as $\lsc[5]{\usf{A}}(\tau)\in\mathcal{M}_{3,1}\pr{\mathbb{R}}$.
The components of $\lsc[5]{\usf{A}}(\tau)$, i.e., $\pr{\lsc[5]{\usf{A}}(\tau)}_{i}, i\in\mathcal{I}$,
for a sequence of time instances are, respectively, shown in subfigures
(a), (b), and (c) in each of Figs.~\ref{fig:sigma0mu1}--\ref{fig:sigma10mu1}.
They are shown using thick gray curves.

Since the predictions of the $\sqrt{\text{AO}}$ and AO algorithms
are derived by, respectively, feeding the $\sqrt{\text{AO}}$ and $\text{AO}$
algorithms the stochastic processes $\lsc{\alpha}_{i}^{\rm Error}$, $\ell\in\mathcal{J}$, they too, in fact, are stochastic processes.
Representative realizations of the predictions from the $\sqrt{\text{AO}}$
(resp. $\text{AO}$) algorithm for different OU parameter sets are,
respectively, shown in Figs.~\ref{fig:sigma0mu1}--\ref{fig:sigma10mu1}
in green (resp. red).

\section{Results and Discussion\label{sec:Results-and-Discussion}}

\subsection{Category I\label{subsec:Category-I}}

\begin{figure}[ht]
\centering{}\includegraphics[width=0.8\textwidth]{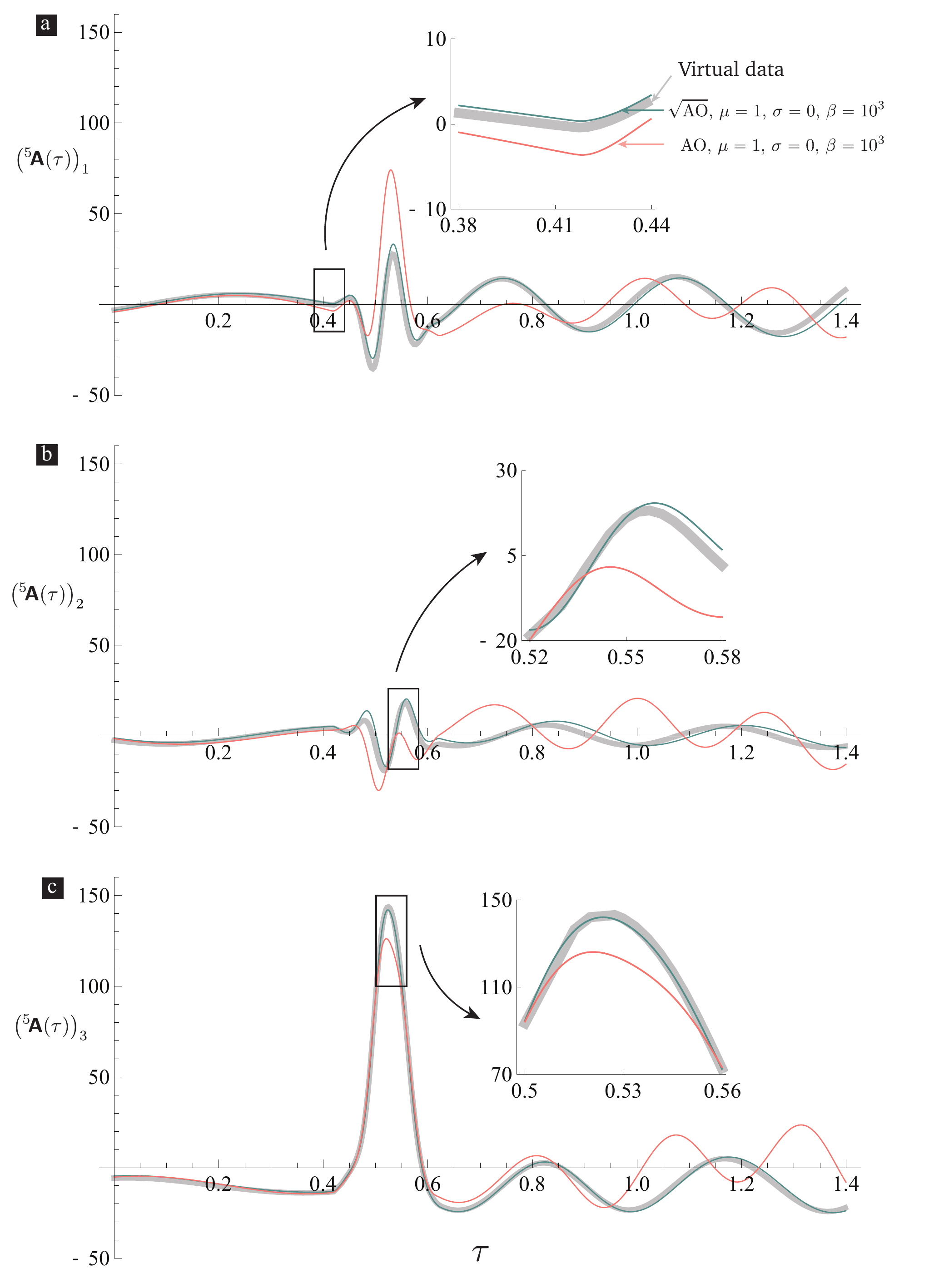}\caption{Comparison of the predictions from the $\sqrt{\text{AO}}$ and $\text{AO}$ algorithms for the acceleration of the material particle $\lsc[5]{\c{X}}$ (see Fig.~\ref{fig:accposition}) in the rigid ellipsoid impact simulation (see \S\ref{subsec:Rigid-ellipsoid-impact} for details). 
Both the $\sqrt{\text{AO}}$ and $\text{AO}$ algorithms were fed the  same virtual error-inclusive accelerometer data. 
The data was generated by adding a particular realization of the OU process to the virtual accelerometer data from the rigid ellipsoid impact simulation. 
The OU realization corresponded to the OU parameter set $(\mu,\sigma,\beta)=(1,0, 10^3)$. 
Subfigures (a), (b), and (c), respectively, show the comparison for the component of $\lsc[5]{\c{X}}$'s acceleration in the $\boldsymbol{e}_{i}$, $i\in \mathcal{I}$, directions. 
}
\label{fig:sigma0mu1}
\end{figure}

Among the OU parameter sets belonging to \ref{category1} the set corresponding to the most amount of error is $\pr{\mu,\sigma}=\pr{1.0, 0.0}$.
Representative realizations of the predictions from the $\sqrt{\text{AO}}$
and AO algorithms for this parameter set are, respectively, shown
in green and red in Fig.~\ref{fig:sigma0mu1}.
In Fig.~\ref{fig:sigma0mu1} the realization of the $\sqrt{\text{AO}}$-algorithm's prediction appears
to be more accurate than that of the AO-algorithm's prediction, especially
with increasing time.
In order to make a more quantitative comparison between the $\sqrt{\text{AO}}$ and AO algorithms' predictions, we focus on the time interval $[0,1]$ and make use of the the metrics

\begin{align}
\epsilon_{2}\pr{\sqrt{\text{AO}}}&:=\frac{\lVert\sqrt{\text{AO}}\left(\lsc[5]\usf{A}\right)-\lsc[5]\usf{A}\rVert_{2}}{\lVert\lsc[5]\usf{A}\rVert_{2}},
\label{equ:L2errorSqrtAO}
\\
\epsilon_{2}\pr{\text{AO}}&:=\frac{\lVert\text{AO}\left(\lsc[5]\usf{A}\right)-\lsc[5]\usf{A}\rVert_{2}}{\lVert\lsc[5]\usf{A}\rVert_{2}},
\label{equ:L2errorAO}
\end{align}
where $\lVert f\rVert_{2}:=\sqrt{\int_{0}^{1}\lVert f(\tau)\rVert^{2}\,d\tau}$;
and $\mathbb{R}\ni\tau\mapsto\sqrt{\text{AO}}\pr{\lsc[5]{\usf{A}}}(\tau)\in\mathcal{M}_{3,1}(\mathbb{R})$, and $\mathbb{R}\ni\tau\mapsto\text{AO}\pr{\lsc[5]{\usf{A}}}(\tau)\in\mathcal{M}_{3,1}(\mathbb{R})$ are, respectively, particular realizations of the $\sqrt{\text{AO}}$ and AO algorithms' predictions for $\mathbb{R}\ni\tau\mapsto\lsc[5]{\usf{A}}(\tau)\in\mathcal{M}_{3,1}(\mathbb{R})$.
The metric $\epsilon_{2}\pr{\sqrt{\text{AO}}}$ (resp. $\epsilon_{2}\pr{\text{AO}}$) is constructed such that the smaller its value the more accurate the realization used in computing it.
The values of $\epsilon_{2}\pr{\sqrt{\text{AO}}}$ and $\epsilon_{2}\pr{\text{AO}}$ for the realizations shown in Fig.~\ref{fig:sigma0mu1} are, respectively, $8.79\%$ and $47.11\%$.
The metric $\epsilon_{2}\pr{\sqrt{\text{AO}}}$'s
smaller value in comparison to that of $\epsilon_{2}\pr{\text{AO}}$
corroborates our earlier assertion that among the $\sqrt{\text{AO}}\pr{\lsc[5]{\usf{A}}}$
and $\sqrt{\text{AO}}\pr{\lsc[5]{\usf{A}}}$ shown in Fig.~\ref{fig:sigma0mu1}
the realization $\sqrt{\text{AO}}\pr{\lsc[5]{\usf{A}}}$ is more accurate.
This comparison between the $\sqrt{\text{AO}}$ and AO algorithms'
predictions' particular realizations prompts us to hypothesize that
the $\sqrt{\text{AO}}$ algorithm is more accurate than the AO algorithm.

In order to compare the $\sqrt{\text{AO}}$ and AO algorithms' predictions
in a more well-balanced and comprehensive manner we calculated $\epsilon_{2}\pr{\sqrt{\text{AO}}}$
and $\epsilon_{2}\pr{\text{AO}}$, respectively, for a large number
of realizations (population size $N=200$) of the predictions from
the $\sqrt{\text{AO}}$ and AO algorithms. The mean values of the thus generated
populations of $\epsilon_{2}\pr{\sqrt{\text{AO}}}$ and $\epsilon_{2}\pr{\text{AO}}$ are $8.788\%$ and $47.114\%$, respectively (see row number 5 of Table.~\ref{Tab:mu_sigma0}). The mean value of the population of $\epsilon_{2}\pr{\sqrt{\text{AO}}}$ being lower than the mean value of the population of $\epsilon_{2}\pr{\text{AO}}$ further supports our earlier hypothesis that $\sqrt{\text{AO}}$-algorithm is more accurate than the AO-algorithm.

To recall, the discussion so far in this section exclusively relates
to the $(\mu,\sigma)$ parameter set $(1.0,0.0)$.
We performed analysis similar to the one discussed in the previous paragraph
for the parameter sets $(0.0, 0.0)$, $(0.1, 0.0)$, $(0.2, 0.0)$, and $(0.5, 0.0)$ as well.
The means of $\epsilon_{2}\pr{\sqrt{\text{AO}}}$'s and $\epsilon_{2}\pr{\text{AO}}$'s populations for these other parameter
sets are, respectively, given in the first and second columns of Table.~\ref{Tab:mu_sigma0}.
It can be seen from Table.~\ref{Tab:mu_sigma0} that the means of the $\epsilon_{2}\pr{\sqrt{\text{AO}}}$ populations are consistently smaller than those of $\epsilon_{2}\pr{\text{AO}}$ populations across all the parameter sets considered.
Furthermore, in Table.~\ref{Tab:mu_sigma0} the difference between the means
of a $\epsilon_{2}\pr{\sqrt{\text{AO}}}$ population and a $\epsilon_{2}\pr{\text{AO}}$ population corresponding to the same parameter set increases with the amount of error, i.e., with the magnitude of $\mu$ in the present category.
Thus for the category of exclusively bias type errors, in addition to the $\sqrt{\text{AO}}$-algorithm appearing to be more accurate than the AO-algorithm, it further appears that the $\sqrt{\text{AO}}$-algorithm's performance over the AO-algorithm increases with increasing amount of error.

\subsection{Category II\label{subsec:Category-II}}
In \ref{category2} we consider the $\pr{\mu,\sigma}$ parameter sets $(0.0, 0.0)$, $(0.0, 1.0)$, $(0.0, 10.0)$, $(0.0, 50.0)$, and $(0.0, 100.0)$.
The means of $\epsilon_{2}\pr{\sqrt{\text{AO}}}$'s and $\epsilon_{2}\pr{\text{AO}}$'s populations for these parameter sets are, respectively, given in the first and second columns of Table.~\ref{Tab:sigma}.
It can be seen from Table.~\ref{Tab:sigma} that the means of the $\epsilon_{2}\pr{\sqrt{\text{AO}}}$ populations are approximately the same as those of $\epsilon_{2}\pr{\text{AO}}$ populations across all the parameter sets considered.
Thus, for the category of exclusively noise type errors the $\sqrt{\text{AO}}$-algorithm appears to perform on par with the AO-algorithm.

\begin{figure}[ht]
\centering{}\includegraphics[width=0.8\textwidth]{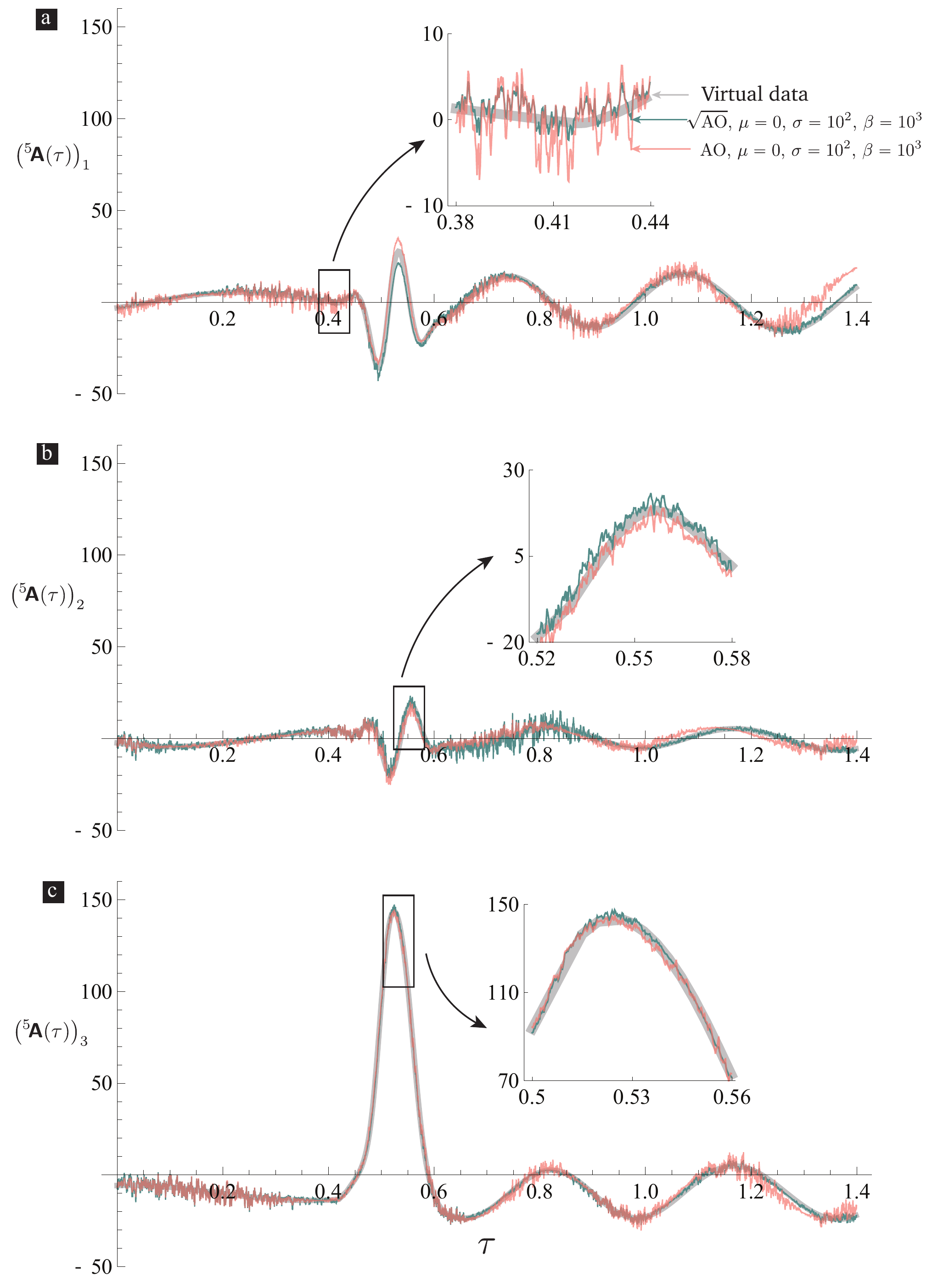}\caption{Comparison of the predictions from the $\sqrt{\text{AO}}$ and $\text{AO}$ algorithms for the acceleration of the material particle $\lsc[5]{\c{X}}$ (see Fig.~\ref{fig:accposition}) in the rigid ellipsoid impact simulation (see \S\ref{subsec:Rigid-ellipsoid-impact} for details). 
Both the $\sqrt{\text{AO}}$ and $\text{AO}$ algorithms were fed the  same virtual error-inclusive accelerometer data. 
The data was generated by adding a particular realization of the OU process to the virtual accelerometer data from the rigid ellipsoid impact simulation. 
The OU realization corresponded to the OU parameter set $(\mu,\sigma,\beta)=(0,10^2, 10^3)$. 
Subfigures (a), (b), and (c), respectively, show the comparison for the component of $\lsc[5]{\c{X}}$'s acceleration in the $\boldsymbol{e}_{i}$, $i\in \mathcal{I}$, directions. 
}
\label{fig:sigma100mu0}
\end{figure}

Among the OU parameter sets belonging to \ref{category2} the set corresponding to the most amount of error is $\pr{\mu,\sigma}=\pr{0.0,100.0}$.
Representative realizations of the predictions from the $\sqrt{\text{AO}}$
and AO algorithms for this parameter set are, respectively, shown
in green and red in Fig.~\ref{fig:sigma100mu0}.
In Fig.~\ref{fig:sigma100mu0}, at least at the earlier time instances, the $\sqrt{\text{AO}}$ and AO algorithms' predictions' realizations are almost indistinguishable from one another.
However, at later time instances the $\sqrt{\text{AO}}$-algorithm seems to be performing better than the AO-algorithm (This feature is likely not reflected
in the results presented in Table.~\ref{Tab:sigma} because they are
calculated only using data from the initial time instances, or, to
be more precise, from the $[0,1]$ time interval.).
Based on this observation we venture to speculate that even when the errors
are predominantly of the noise type, the $\sqrt{\text{AO}}$-algorithm will eventually begin to outperform the AO-algorithm.

\subsection{Category III}

\begin{figure}[ht]
\centering{}\includegraphics[width=0.8\textwidth]{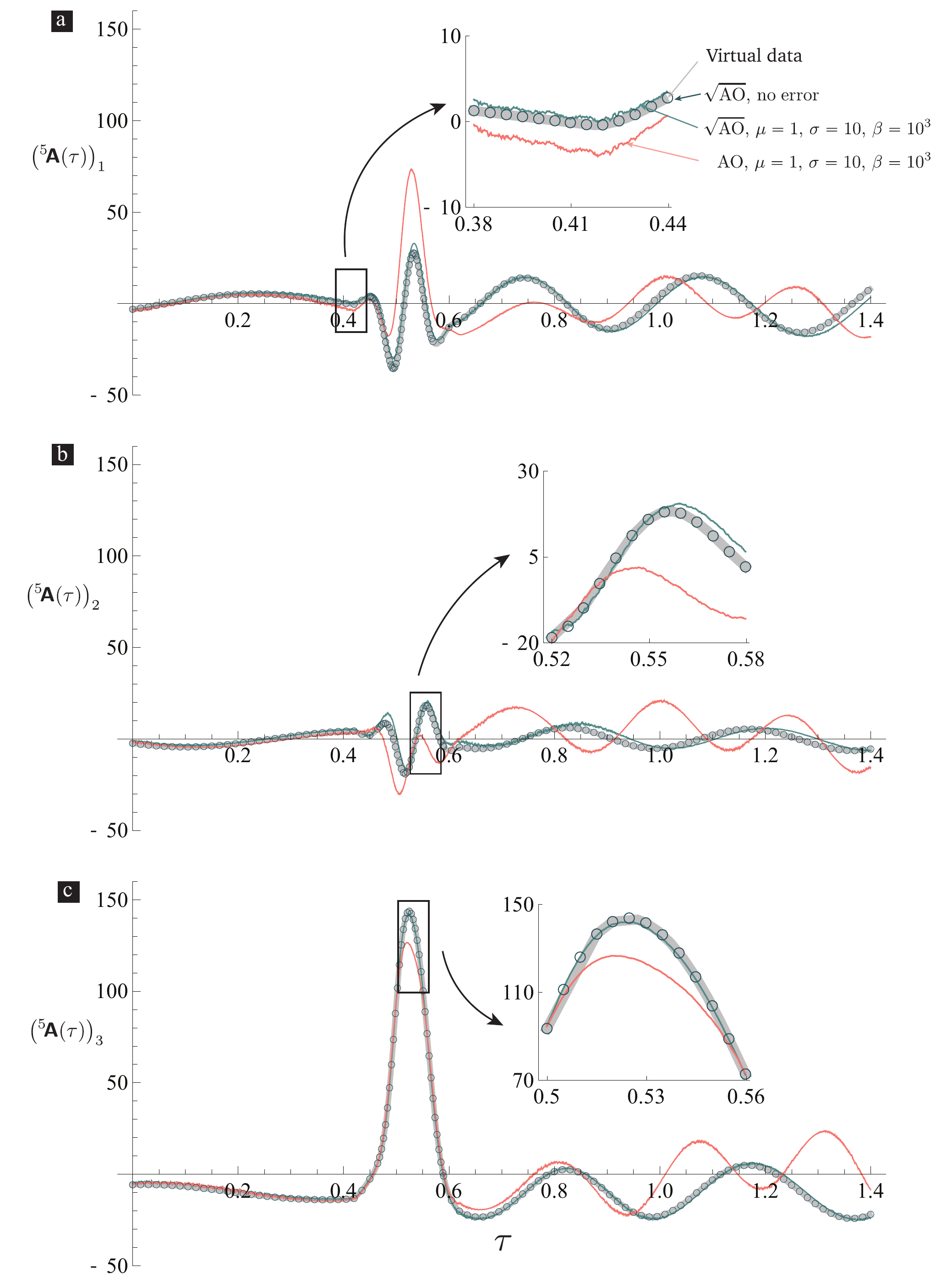}\caption{Comparison of the predictions from the $\sqrt{\text{AO}}$ and $\text{AO}$ algorithms for the acceleration of the material particle $\lsc[5]{\c{X}}$ (see Fig.~\ref{fig:accposition}) in the rigid ellipsoid impact simulation (see \S\ref{subsec:Rigid-ellipsoid-impact} for details). 
Both the $\sqrt{\text{AO}}$ and $\text{AO}$ algorithms were fed the  same virtual error-inclusive accelerometer data. 
The data was generated by adding a particular realization of the OU process to the virtual accelerometer data from the rigid ellipsoid impact simulation. 
The OU realization corresponded to the OU parameter set $(\mu,\sigma,\beta)=(1,10, 10^3)$. 
Subfigures (a), (b), and (c), respectively, show the comparison for the component of $\lsc[5]{\c{X}}$'s acceleration in the $\boldsymbol{e}_{i}$, $i\in \mathcal{I}$, directions. 
The predictions of the $\sqrt{\text{AO}}$ algorithm when fed just the virtual accelerometer data, i.e., with no added errors, is also shown in (a), (b), and (c) using black open circles.  
}
\label{fig:sigma10mu1}
\end{figure}

In \ref{category3} we consider the $\pr{\mu,\sigma}$ parameter
sets $(0.0, 10.0)$, $(0.1, 10.0)$, $(0.2, 10.0)$, $(0.5, 10.0)$, and $(1, 10.0)$.
The means of $\epsilon_{2}\pr{\sqrt{\text{AO}}}$'s and $\epsilon_{2}\pr{\text{AO}}$'s populations for these parameter sets are, respectively, given in the first and second columns of Table.~\ref{Tab:mu}.
Among the OU parameter sets belonging to \ref{category3} the
set corresponding to the most amount of error is $\pr{\mu,\sigma}=\pr{1.0,10.0}$.
Representative realizations of the predictions from the $\sqrt{\text{AO}}$
and AO algorithms for this parameter set are, respectively, shown
in green and red in Fig.~\ref{fig:sigma10mu1}.

It can be seen from Table.~\ref{Tab:mu} that the means
of the $\epsilon_{2}\pr{\sqrt{\text{AO}}}$ populations are consistently
smaller than those of $\epsilon_{2}\pr{\text{AO}}$ populations across
all the parameter sets considered.
Furthermore, in Table.~\ref{Tab:mu} the difference between the means
of a $\epsilon_{2}\pr{\sqrt{\text{AO}}}$ population and a $\epsilon_{2}\pr{\text{AO}}$ population corresponding to the same parameter set increases with the amount of error, i.e., with the magnitudes of $\sigma$ and $\mu$.
Thus, in \ref{category3} the relative performance of the $\sqrt{\text{AO}}$
and AO algorithms is very similar to that in \ref{category1}.

From the discussion in $\S$\ref{subsec:Category-I} we know that the AO-algorithm is more sensitive to bias type errors than the $\sqrt{\text{AO}}$-algorithm and from the discussion in $\S$\ref{subsec:Category-II} we know that the $\sqrt{\text{AO}}$ and AO algorithms are, approximately, equally sensitive to noise type errors.
From the results in this section it appears that the $\sqrt{\text{AO}}$-algorithm outperforms the AO-algorithm as long as the errors have some bias type component in them, irrespective of what the amount of the noise type component in them is.

\begingroup
\setlength{\tabcolsep}{10pt}
\renewcommand{\arraystretch}{1.5}

\begin{table}[ht!]
\centering\caption{The mean and standard deviation of the error measure $\epsilon_{2}$
for 200 realizations of the accelerometer data only containing bias
type error with $\sigma=0$. In this case, as the value of standard
deviation is quite small compared to the value of mean, we do not
show the value in the table.\label{Tab:mu_sigma0}}
\begin{tabular}{|>{\columncolor[gray]{0.9}}c|l|l|}
\hline
\cellcolor[gray]{0.7} & \multicolumn{2}{c|}{\cellcolor[gray]{0.7}$\epsilon_{2}\times10^{3}$ (mean$\pm$ std)}\tabularnewline
\hline
\multirow{1}{*}{\cellcolor[gray]{0.7}$\mu$} & \cellcolor[gray]{0.9}$\sqrt{\text{AO}}$-algorithm & \cellcolor[gray]{0.9} AO-algorithm\tabularnewline
\hline
$0$ & $0.01$ & $0.02$\tabularnewline
\hline
$0.1$ & $8.87$ & $41.02$\tabularnewline
\hline
$0.2$ & $17.73$ & $83.63$\tabularnewline
\hline
$0.5$ & $44.22$ & $220.28$\tabularnewline
\hline
$1$ & $87.88$ & $471.14$\tabularnewline
\hline
\end{tabular}
\end{table}

\endgroup

\begingroup
\setlength{\tabcolsep}{10pt}
\renewcommand{\arraystretch}{1.5}

\begin{table}[ht!]
\centering\caption{The mean and standard deviation of the error measure $\epsilon_{2}$
for 200 realizations of the accelerometer data only containing noise
type error with $\mu=0$\label{Tab:sigma}}
\begin{tabular}{|>{\columncolor[gray]{0.9}}c|l|l|}
\hline
\cellcolor[gray]{0.7} & \multicolumn{2}{c|}{\cellcolor[gray]{0.7}$\epsilon_{2}\times10^{3}$ (mean$\pm$ std)}\tabularnewline
\hline
\multirow{1}{*}{\cellcolor[gray]{0.7}$\sigma$} & \cellcolor[gray]{0.9}$\sqrt{\text{AO}}$-algorithm & \cellcolor[gray]{0.9} AO-algorithm\tabularnewline
\hline
0 & $0.01$ & $0.02$\tabularnewline
\hline
$1$ & $1.03\pm0.09$ & $1.16\pm0.21$\tabularnewline
\hline
$10$ & $10.22\pm0.64$ & $11.46\pm1.80$\tabularnewline
\hline
$50$ & $52.79\pm3.33$ & $56.87\pm7.79$\tabularnewline
\hline
$100$ & $115.85\pm10.11$ & $112.83\pm15.54$\tabularnewline
\hline
\end{tabular}
\end{table}

\endgroup

\begingroup
\setlength{\tabcolsep}{10pt}
\renewcommand{\arraystretch}{1.5}

\begin{table}[ht!]
\centering\caption{The mean and standard deviation of the error measure $\epsilon_{2}$
for 200 realizations of the accelerometer data containing bias and
noise type errors with $\sigma=10$\label{Tab:mu}}
\begin{tabular}{|>{\columncolor[gray]{0.9}}c|l|l|}
\hline
\cellcolor[gray]{0.7} & \multicolumn{2}{c|}{\cellcolor[gray]{0.7}$\epsilon_{2}\times10^{3}$ (mean$\pm$ std)}\tabularnewline
\hline
\multirow{1}{*}{\cellcolor[gray]{0.7}$\mu$} & \cellcolor[gray]{0.9}$\sqrt{\text{AO}}$-algorithm & \cellcolor[gray]{0.9} AO-algorithm\tabularnewline
\hline
$0$ & $10.22\pm0.64$ & $11.46\pm1.80$\tabularnewline
\hline
$0.1$ & $13.42\pm1.14$ & $42.61\pm5.05$\tabularnewline
\hline
$0.2$ & $20.19\pm1.43$ & $84.52\pm5.57$\tabularnewline
\hline
$0.5$ & $45.18\pm1.47$ & $221.08\pm6.02$\tabularnewline
\hline
$1$ & $88.27\pm1.44$ & $472.26\pm6.47$\tabularnewline
\hline
\end{tabular}
\end{table}

\endgroup

\clearpage{}

\section{Concluding remarks\label{Sec:Con}}
\begin{enumerate}
\item The results discussed in $\S$\ref{sec:Results-and-Discussion} show
that the $\sqrt{\text{AO}}$-algorithm provides a valid approach to
determine the complete motion of a rigid body using only data from
four tri-axial accelerometers.
However, the $\sqrt{\text{AO}}$-algorithm's practical validity in the field still remains to be explored.
In the future, we plan to conduct an experimental evaluation of the $\sqrt{\text{AO}}$-algorithm to compliment its \textit{in silico} validation that we presented in this paper.
\item The comparison in $\S$\ref{sec:Results-and-Discussion} shows that
for the cases we considered the $\sqrt{\text{AO}}$-algorithm is less
sensitive to bias type errors compared to the AO-Algorithm.
However, we have not provided a mathematical proof that the $\sqrt{\text{AO}}$-algorithm is better than the AO-algorithm with regard to bias type errors.
Thus, though the comparison presented in $\S$\ref{sec:Results-and-Discussion} provides strong support to the hypothesis that the $\sqrt{\text{AO}}$-algorithm
is less sensitive to bias type errors than the AO-algorithm, it by
no means provides a proof for the hypothesis.
A definitive resolution to the question of whether the hypothesis is true requires an error analysis of both the $\sqrt{\text{AO}}$-algorithm as well as the AO-algorithm.
We currently have not carried out such analyses. Nevertheless, irrespective
of the relative merit of the $\sqrt{\text{AO}}$ over the AO algorithm, it is quite clear from its derivation and the results discussed in $\S$\ref{sec:Results-and-Discussion} that it provides a valid approach
for determining the complete motion of a rigid body from accelerometer
data.
\item The $\sqrt{\text{AO}}$-algorithm retains all the benefits of the
AO-algorithm. Both algorithms provide the complete motion of the rigid
body in the fixed laboratory frame. Without integration or differentiation,
both algorithms are able to determine the pseudo acceleration field,
providing the magnitude of acceleration for all material particles.
Both algorithms can be applied to any arrangement of four tri-axial
accelerometers as long as they do not lie in the same plane. There
is no restriction on the orientation of the tri-axial accelerometers.
\item In the \textit{in silico} validation, evaluation, and comparison of
the $\sqrt{\text{AO}}$-algorithm that we set up in $\S$\ref{sec:In-silico-validation} we used the OU process to model experimental errors.
Even more specifically, we took the magnitude of the parameter $\mu$ in the OU process as a measure of the bias type errors in the OU process' realizations.
There of course exist bias type errors that cannot be modeled in this
manner. Thus, our evaluation of the relative sensitivities of the
$\sqrt{\text{AO}}$ and the AO algorithms to bias type errors was
carried out using a limited form of bias type errors. A more general
method to represent bias type errors in the virtual accelerometer
data would provide a more comprehensive comparison of the relative
sensitivities of the $\sqrt{\text{AO}}$ and the AO algorithms to
bias type errors.
\item We expect the $\sqrt{\text{AO}}$-algorithm to be especially useful
for constructing inputs to the upcoming finite element based brain
injury criteria. The finite element based brain injury criteria are
based on the mechanics of head motion and brain deformation, while
the traditional brain injury criteria have mostly been developed empirically.
Therefore, we expect the finite element based brain injury criteria
to find increased use in the future.
\end{enumerate}

\section*{Funding information}

The authors gratefully acknowledge support from the Panther Program
and the Office of Naval Research (Dr. Timothy Bentley) under grants
N000141812494 and N000142112044.

\section*{Declaration of Competing Interest}

The authors declare that they have no known competing financial interests
or personal relationships that could have appeared to influence the
work reported in this paper.

\section*{Acknowledgments}

The authors thank Sayaka Kochiyama for her help in preparing some
of the figures in the manuscript.

\appendix
\section{Definition of the map $\star\pr{\cdot}$\label{subsec:hodgestar}}

For $\nsd=2$ , the map $\star\pr{\cdot}:\mathfrak{so}(\mathbb{R},2)\to\m{R}$
is defined by the equation $\star\pr{\cdot}=\pr{\cdot}_{21}$. The
inverse of $\star\pr{\cdot}$ is the map $\star^{-1}\pr{\cdot}:\m{R}\to\mathfrak{so}(\mathbb{R},2)$
defined by the equation

\begin{align}
\star\pr{\alpha}=\begin{pmatrix}
0 & -\alpha\\
\alpha & 0
\end{pmatrix}.
\label{equ:hodge2d}
\end{align}

For $\nsd=3$ , the map $\star\pr{\cdot}:\mathfrak{so}(\mathbb{R},3)\to\mathcal{M}_{3,1}(\mathbb{R})$
is defined by the equation $\star\pr{\cdot}=\pr{\pr{\cdot}_{32},\pr{\cdot}_{13},\pr{\cdot}_{21}}$.
The inverse of $\star\pr{\cdot}$ is the map$ \star^{-1}\pr{\cdot}:\mathcal{M}_{3,1}(\mathbb{R})\to\mathfrak{so}(\mathbb{R},3)$
defined by the equation

\begin{align}
\star\pr{\pr{\alpha_1,\alpha_2,\alpha_3}}=\begin{pmatrix}
0 & -\alpha_3 &\alpha_2\\
\alpha_3 & 0 & -\alpha_1\\
-\alpha_2 &\alpha_1 &0
\end{pmatrix}.
\label{equ:hodge3d}
\end{align}

To make our notation appear less cumbersome we denote $\star^{-1}\pr{\cdot}$
too as $\star\pr{\cdot}$. Whether we mean $\star\pr{\cdot}$ or $\star^{-1}\pr{\cdot}$ will be clear from the argument of $\star\pr{\cdot}$.

\section{Derivation of $~\eqref{equ:sym-1}$, i.e., proof of the statement
that square of $\busf{W}(\tau)$ is equal to the symmetric part of $\usf{P}(\tau)$}
\label{subsec:square}

The following lemmas can be shown to be equivalent to some of the standard
results in the mechanics of rigid solids, see the work in \cite[\S{2.5.2}]{jog2015} and \cite[\S{6.4}]{oreilly2020}, which treats the rigid body motion in a modern continuum mechanics style; or see the work in \cite[\S{9.4}]{maruskin2018} and \cite[\S{15}]{marsden2013}, which treats the rigid body motion from a perspective of geometric mechanics.
However, at a cursory level, due to our notation and formalism, those results might appear to be different from the below lemmas.
The differences in notation and formalism are primarily due to the
fact that in our work we distinguish between the vector spaces to
which the various physical quantities, e.g., the rotation operation,
belong to and the (non-dimensional) matrix vector spaces to which
the component representations of those quantities belong to.
For that reason, we believed that it would be helpful to the reader
if we presented the following lemmas using the notation and formalism that we use in the current work.

\subsection{Skew symmetry of $\busf{W}(\tau)$}

\begin{lemma}

The matrix $\busf{W}(\tau)$, defined in $\eqref{eq:busfW}$, is skew-symmetric.
\label{lemma:skew}

\end{lemma}

\textit{Proof}. It can be shown using $\busf{W}(\tau)$'s definition $\eqref{eq:busfW}$ and equations $\eqref{equ:wdeNDF}$ and $\eqref{equ:QQ:1}$ that
\begin{equation}
\busf{W}(\tau)=\tpsb{\usf{Q}}(\tau)\,\usf{Q}'(\tau).
\label{eq:WbeqQtQp}
\end{equation}

Differentiating $\eqref{equ:QQ:1}$ we
get
\begin{equation}
\left(\usf{Q}^{\sf T}\right)'(\tau)\,\usf{Q}(\tau)+\tpsb{\usf{Q}}(\tau)\,\usf{Q}'(\tau)=\usf{0}.\label{equ:QQ0}
\end{equation}
Noting that $\pr{\tpsb{\usf{Q}}}'(\tau)=\tps{\usf{Q}'}(\tau)$ we see
that the first term on the left hand side of \eqref{equ:QQ0} is equal
to $\tps{\usf{Q}'}(\tau)\,\usf{Q}(\tau)$, which, in fact, is equal
to the transpose of the second term on the left hand side of \eqref{equ:QQ0}. Thus, it follows from \eqref{equ:QQ0} that $\text{sym}\pr{\tpsb{\usf{Q}}(\tau)\,\usf{Q}'(\tau)}=\usf{0}$.
That is, that $\tpsb{\usf{Q}}(\tau)\,\usf{Q}'(\tau)$ is skew-symmetric.
The result that $\busf{W}(\tau)$ too is skew-symmetric now immediately
follows from $\eqref{eq:WbeqQtQp}$.

\subsection{Derivation of equation $~\eqref{equ:sym-1}$\label{theorem:skeww}}

\begin{lemma}

The symmetric part of $\usf{P}(\tau)$ is equal to the square of $\busf{W}(\tau)$.

\end{lemma}

\textit{Proof}. Differentiating $\eqref{equ:QQ:1}$ twice and rearranging
we get that

\begin{align}
\left(\usf{Q}^{\sf T}\right)''(\tau)\,\usf{Q}(\tau)+\tpsb{\usf{Q}}(\tau)\,\usf{Q}''(\tau)&=-2\left(\usf{Q}^{\sf T}\right)'(\tau)\,\usf{Q}'(\tau).
\label{equ:diffQ}
\end{align}

The equation $\eqref{equ:diffQ}$ on noting that the first and second terms on its left hand side are in fact transposes of each other simplifies to

\begin{align}
\text{sym}\pr{\tpsb{\usf{Q}}(\tau)\,\usf{Q}''(\tau)}&=-\left(\usf{Q}^{\sf T}\right)'(\tau)\,\usf{Q}'(\tau).
\label{equ:diffQ2}
\end{align}

 Writing the term on the right hand side
of $\eqref{equ:diffQ2}$ as $-\left(\usf{Q}^{\sf T}\right)'(\tau)\,\usf{I}\,\usf{Q}'(\tau)$,
and then using $\eqref{equ:QQ:2}$ and replacing the $\usf{I}$ in the
resulting equation with $\usf{Q}(\tau)\tpsb{\usf{Q}}(\tau)$, we get

\begin{equation}
\text{sym}\pr{\tpsb{\usf{Q}}(\tau)\,\usf{Q}''(\tau)}=-
\pr{\left(\usf{Q}^{\sf T}\right)'(\tau)\,\usf{Q}(\tau)}\,\pr{\tpsb{\usf{Q}}(\tau)\,\usf{Q}'(\tau)}.
\label{equ:diffQ3}
\end{equation}
Noting that $\pr{\tpsb{\usf{Q}}}'(\tau)=\tps{\usf{Q}'}(\tau)$
we see that the first factor on the right hand side of $\eqref{equ:diffQ3}$
is equal to $\tps{\usf{Q}'}(\tau)\,\usf{Q}(\tau)$, which is the transpose
of the second factor on the right hand side of $\eqref{equ:diffQ3}$,
namely $\tpsb{\usf{Q}}(\tau)\,\usf{Q}'(\tau)$.
We, however, know from $\eqref{eq:WbeqQtQp}$ that this second factor
is equal to $\busf{W}(\tau)$. Thus, we get from $\eqref{equ:diffQ3}$
that

\begin{subequations}
\begin{align}
\text{sym}\pr{\tpsb{\usf{Q}}(\tau)\,\usf{Q}''(\tau)}&=
-\tpsb{\busf{W}}(\tau)\,\busf{W}(\tau),
\label{eq:PW1}
\intertext{which simplifies on using Lemma \ref{lemma:skew} to}
\text{sym}\pr{\tpsb{\usf{Q}}(\tau)\,\usf{Q}''(\tau)}&=
\busf{W}^{2}(\tau).
\label{eq:PW2}
\end{align}
\label{eq:PW}
\end{subequations}

Using $\eqref{equ:pde}$ and replacing the quantity $\tpsb{\usf{Q}}(\tau)\,\usf{Q}''(\tau)$
appearing on the left hand side of $\eqref{eq:PW2}$ with $\usf{P}(\tau)$, we get
\begin{equation}
\text{sym}\left(\usf{P}(\tau)\right)=\busf{W}^{2}(\tau).
\end{equation}

\section{The matrix $\text{sym}\pr{\usf{P}(\tau)}$ is negative semidefinite and its negative eigenvalues, if they exist, have even algebraic multiplicities}
\label{subsec:sqrt_prop}

The entries of $\busf{W}(\tau)$ and $\text{sym}\pr{\usf{P}(\tau)}$
are all real numbers. However, in this section we consider $\busf{W}(\tau)$
and $\text{sym}\pr{\usf{P}(\tau)}$ to be elements of $\mathcal{M}_{\nsd,\nsd}(\mathbb{C})$
, where $\mathcal{M}_{\nsd,\nsd}(\mathbb{C})$ is the space of all $\nsd\times\nsd$ matrices whose entries belong to $\mathbb{C}$, the space of complex numbers.

\subsection{Negative semi-definiteness of $\text{sym}\pr{\usf{P}(\tau)}$}

\begin{lemma}
The matrix $\text{sym}\pr{\usf{P}(\tau)}$ is negative semidefinite.
\label{lemma:negative}
\end{lemma}

\textit{Proof}. The matrix $\text{sym}\pr{\usf{P}(\tau)}$ is self-adjoint
since it is equal to its transpose, which, as all of $\text{sym}\pr{\usf{P}(\tau)}$'s
entries are real, is equal to its conjugate-transpose, i.e., to
its adjoint. Hence it follows from \cite[7.31]{Axler2015}
that the matrix $\text{sym}\pr{\usf{P}(\tau)}$ is negative semidefinite
iff the inner product $\langle\text{sym}\pr{\usf{P}(\tau)}\usf{X},\usf{X}\rangle$,
where $\usf{X}\in\mathcal{M}_{\nsd,1}(\mathbb{C})$ but is otherwise
arbitrary, is always non-positive.

It follows from $\eqref{equ:sym-1}$ that \begin{equation}
\langle\text{sym}\pr{\usf{P}(\tau)}\usf{X},\usf{X}\rangle=\langle\busf{W}^2(\tau)\usf{X},\usf{X}\rangle.
\label{equ:sym1xxt}
\end{equation}
Since we know from Lemma \ref{lemma:skew} that $\busf{W}(\tau)$ is skew-symmetric, we can write $\busf{W}^2(\tau)$ on the right hand side of $\eqref{equ:sym1xxt}$ as $-\tpsb{\busf{W}}(\tau)\,\busf{W}(\tau)$.
On doing so and using the properties of the inner product we get \begin{equation}
\langle\text{sym}\pr{\usf{P}(\tau)}\usf{X},\usf{X}\rangle=-\langle\busf{W}(\tau)\usf{X},\busf{W}(\tau)\usf{X}\rangle.
\label{equ:W2eqngWtW}
\end{equation}
It also follows from the properties of the inner product that $\langle\busf{W}(\tau)\usf{X},\busf{W}(\tau)\usf{X}\rangle$ is always non-negative. Therefore we get from  $\eqref{equ:W2eqngWtW}$ that $\langle\text{sym}\pr{\usf{P}(\tau)}\usf{X},\usf{X}\rangle$ is always non-positive, or equivalently that $\text{sym}\pr{\usf{P}(\tau)}$ is negative semidefinite.

\subsection{Form of the eigenvalues of $\text{sym}\pr{\usf{P}(\tau)}$}

\begin{lemma}

The matrix $\text{sym}\pr{\usf{P}(\tau)}$'s negative eigenvalues, if they exist, have even algebraic multiplicities.
\label{lemma:form}

\end{lemma}

\textit{Proof.} Say $\usf{S}\in\mathcal{M}_{\nsd,\nsd}(\mathbb{C})$
then $\usf{S}$ is said to be normal when it commutes with its conjugate-transpose $\hpsb{\usf{S}}$, i.e., when $\usf{S}\hpsb{\usf{S}}=\hpsb{\usf{S}}\usf{S}$. Note that
\begin{equation}
\busf{W}(\tau)\tpsb{\busf{W}}(\tau)=\busf{W}(\tau)\pr{-\busf{W}(\tau)}=\pr{-\busf{W}(\tau)}\busf{W}(\tau)=\tpsb{\busf{W}}(\tau)\busf{W}(\tau).
\label{equ:WbisNormal}
\end{equation}
The first and the third equalities in $\eqref{equ:WbisNormal}$ follow
from the fact that $\busf{W}(\tau)$ is skew-symmetric (Lemma \ref{lemma:skew}). Since all the entries of $\busf{W}(\tau)$ are real, the transpose of $\busf{W}(\tau)$ is equal to its conjugate-transpose. For this reason it follows from $\eqref{equ:WbisNormal}$ that $\busf{W}(\tau)\hpsb{\busf{W}}(\tau)=\hpsb{\busf{W}}(\tau)\busf{W}(\tau)$, or equivalently that $\busf{W}(\tau)$ is normal.

Since $\busf{W}(\tau)$ is normal it follows from the \textit{Complex
Spectral Theorem} e.g., see, \cite[ 7.24]{Axler2015} that there exists
an unitary matrix $\usf{U}(\tau)\in\mathcal{M}_{\nsd,\nsd}(\mathbb{C})$
such that $\textrm{\ensuremath{\hpsb{\usf{U}}(\tau)\busf{W}(\tau)\usf{U}(\tau)} }=\text{diag}\pr{\mu_i(\tau)}_{i\in\mathcal{I}}$, where $\mu_i(\tau)\in\mathbb{C}$ and $\text{diag}\pr{\mu_i(\tau)}_{i\in\mathcal{I}}$ is a diagonal matrix whose diagonal entries are $\mu_1(\tau),\mu_2(\tau)\ldots,\mu_{\nsd}(\tau)$. That is,

\begin{equation}
\textrm{\ensuremath{\hpsb{\usf{U}}(\tau)\busf{W}(\tau)\usf{U}(\tau)} }=
\begin{pmatrix}\mu_{1}(\tau) & 0 &\dots & 0\\0 &\mu_{2}(\tau) &\dots & 0\\\vdots &\vdots &\ddots &\vdots\\0 & 0 &\dots &\mu_{\nsd}(\tau)\end{pmatrix}.
\label{equ:UhWbUeqDiagMus}
\end{equation}
The complex numbers $\mu_i(\tau)$, not necessarily distinct, are the eigenvalues of $\busf{W}(\tau)$ and the columns of $\usf{U}(\tau)$ are the eigenvectors of $\busf{W}(\tau)$.
To be more specific,\begin{equation}
\busf{W}(\tau)\usf{u}_i(\tau)=\mu_i(\tau)\usf{u}_i(\tau)~\text{(no sum over $i$)},
\label{equ:Wbuieqmuiui}
\end{equation}
where $\usf{u}_i(\tau) =\pr{\pr{\usf{U}(\tau)}_{ji}}_{j\in\mathcal{I}}$.
Applying the operation of complex-conjugation to both sides of $\eqref{equ:Wbuieqmuiui}$ and noting that the entries of $\busf{W}(\tau)$ are all real we get that
\begin{align}
\busf{W}(\tau)\usf{u}_i^*(\tau)
&=\mu_i^*(\tau)\,\usf{u}_i^*(\tau)
~\text{(no sum over $i$)},
\label{equ:WbuieqmuiuiCC}
\end{align}
where $\mu_i^*(\tau)$ and $\usf{u}_i^{\ast}(\tau)$ are, respectively, the complex-conjugates of $\mu_i(\tau)$ and $\usf{u}_i(\tau)$.
\sloppy It follows from $\eqref{equ:WbuieqmuiuiCC}$ that if $\mu_i(\tau)$ is an eigenvalue of $\busf{W}(\tau)$ then so is $\mu_i^*(\tau)$.
Thus $\pr{\mu_i(\tau)}_{i\in\mathcal{I}}$ has the form $\varsigma\pr{z_{1}(\tau), z_1^*(\tau), z_2(\tau), z_2^*(\tau),\ldots, z_k(\tau), z_k^*(\tau),\alpha_1(\tau),\alpha_2(\tau),\ldots,\alpha_l(\tau)}$, where $\varsigma(\cdot)$ is the permutation operation, $z_i(\tau)\in\m{C}$ with $\text{Im}\pr{z_i(\tau)}\neq 0$, $z_i^*(\tau)$ is the complex-conjugate
of $z_i(\tau)$, $0\le k\le\left\lfloor\nsd/2\right\rfloor$\footnote{Here $\left\lfloor\cdot\right\rfloor$ is the \textit{floor} function.}, $\alpha_i(\tau)\in\m{R}$, and $l=\nsd-2 k$.
It is not necessary that the complex numbers $z_i(\tau)$ be distinct from one another.
The same is the case with the real numbers $\alpha_i(\tau)$.

Taking the square on both sides of $\eqref{equ:UhWbUeqDiagMus}$ and
using our knowledge about the form of $\pr{\mu_i(\tau)}_{i\in\mathcal{I}}$
we get that
\begin{equation}
\pr{\textrm{\ensuremath{\hpsb{\usf{U}}(\tau)\busf{W}(\tau)\usf{U}(\tau)} }}^2=\text{diag}\,\varsigma\pr{\prb{z_{1}}^2(\tau),\prb{z_1^*}^2(\tau),\ldots,\prb{z_{k}}^2(\tau),\prb{z_{k}^*}^2(\tau),\alpha_1^2(\tau),\ldots,\alpha_l^2(\tau)}.\label{equ:UhWbUeqDiagMusSq}
\end{equation}
The expression on the left hand side of $\eqref{equ:UhWbUeqDiagMusSq}$
can be simplified as $\pr{\textrm{\ensuremath{\hpsb{\usf{U}}(\tau)\busf{W}(\tau)\usf{U}(\tau)} }}^2=\hpsb{\usf{U}}(\tau)\busf{W}(\tau)\usf{U}(\tau)\hpsb{\usf{U}}(\tau)\busf{W}(\tau)\usf{U}(\tau)=\hpsb{\usf{U}}(\tau)\busf{W}^2(\tau)\usf{U}(\tau)=\hpsb{\usf{U}}(\tau)\text{sym}\pr{\usf{P}(\tau)}\usf{U}(\tau)$,
where the second equality follows from the fact that $\usf{U}(\tau)$
is an unitary matrix, and the third equality from \eqref{equ:sym-1}.
Thus we can get from$\eqref{equ:UhWbUeqDiagMusSq}$ that
\begin{subequations}
\begin{align}
\hpsb{\usf{U}}(\tau)\text{sym}\pr{\usf{P}(\tau)}\usf{U}(\tau)&=\text{diag}\,\varsigma\pr{\prb{z_{1}}^2(\tau) ,\prb{z_1^*}^2(\tau),\ldots,\prb{z_{k}}^2(\tau),\prb{z_{k}^*}^2(\tau),\alpha_1^2(\tau),\ldots,\alpha_l^2(\tau)},
\label{eq:FormofD}
\end{align}
\label{equ:DeCompOfSymP}
\end{subequations}
which implies that $\usf{u}_i(\tau)$ are the eigenvectors of $\text{sym}\pr{\usf{P}(\tau)}$ as well, but with their corresponding eigenvalues being $\varsigma\pr{\prb{z_{1}}^2(\tau),\prb{z_1^*}^2(\tau),\ldots,\prb{z_{k}}^2(\tau),\prb{z_{k}^*}^2(\tau),\alpha_1^2(\tau),\ldots,\alpha_l^2(\tau)}$.
We know from Lemma \ref{lemma:negative} that all of $\text{sym}\pr{\usf{P}(\tau)}$'s eigenvalues are non-positive. Therefore, each $\prb{z_i}(\tau)$ must be of the form $\lambda_i(\tau)\sqrt{-1}$,
where $\lambda_i(\tau)\in\mathbb{R}$ and $\lambda_i(\tau)\neq0$, and $\alpha_i(\tau)=0$. Hence, we get from $\eqref{eq:FormofD}$ that

\begin{equation}
\hpsb{\usf{U}}(\tau)\text{sym}\pr{\usf{P}(\tau)}\usf{U}(\tau)=\text{diag}\,\varsigma\pr{-\lambda_1^2(\tau) ,-\lambda_1^2(\tau),\ldots, -\lambda_k^2(\tau) ,-\lambda_k^2(\tau), 0,\ldots,0},
\label{equ:ComplexPsymDecomposition}
\end{equation}
where, to reiterate, $0\le k\le\left\lfloor\nsd/2\right\rfloor$ and $\lambda_i(\tau)$, when they exist, are non-zero and not necessarily distinct.
It can be noted from this last assertion that all of $\text{sym}\pr{\usf{P}(\tau)}$'s negative eigenvalues, specifically those corresponding to $\lambda_i(\tau)$, are of even geometric multiplicities.
For the case of symmetric matrices, algebraic and geometric multiplicities
are one and the same. Therefore, $\text{sym}\pr{\usf{P}(\tau)}$'s
negative eigenvalues, when they exist, are also of even algebraic multiplicities.

\section{Calculating $\busf{W}(\tau)$ as the square-root of $\text{sym}\pr{\usf{P}(\tau)}$\label{subsec:sqrt_procedure}}

\subsection{A spectral decomposition of $\text{sym}\pr{\usf{P}(\tau)}$\label{subsec:A-spectral-decomposition}}

Since $\text{sym}\pr{\usf{P}(\tau)}$ is a real symmetric matrix it
follows from the \textit{Real Spectral Theorem} \cite[7.29]{Axler2015} that it can be decomposed as
\begin{equation}
\usf{N}(\tau)\,\usf{D}(\tau)\,\tpsb{\usf{N}}(\tau),
\label{eq:RealSpectralDecompositionofSymP}
\end{equation}
where $\usf{D}(\tau)$ and $\usf{N}(\tau)$ belong to $\mathcal{M}_{\nsd,\nsd}(\mathbb{R})$.
We first describe $\usf{N}(\tau)$ and then $\usf{D}(\tau)$, in the next paragraph.
The matrix $\usf{N}(\tau):=\tps{\usf{n}_1(\tau),\ldots,\usf{n}_{\nsd}(\tau)}$ where $\usf{n}_i(\tau)\in\mathcal{M}_{\nsd,1}(\mathbb{R})$ are $\text{sym}\pr{\usf{P}(\tau)}$'s eigenvectors that are constructed such that $\usf{n}_i(\tau)\cdot\usf{n}_{j}(\tau)=\delta_{ij}$, or equivalently
\begin{equation}
\usf{N}(\tau)\tpsb{\usf{N}}(\tau)=\usf{I}.
\label{eq:NOrthonormalityCondition}
\end{equation}

Using $\eqref{equ:ComplexPsymDecomposition}$ it can be shown that for $\nsd=1$, $\usf{D}(\tau)=\pr{0}$; for $\nsd=2$, $\usf{D}(\tau)=\text{diag}\pr{0,0}$ or $\text{diag}\pr{-\lambda_1^2(\tau),-\lambda_1^2(\tau)}$, where $\lambda_1(\tau)\neq0$; and for $\nsd=3$, $\usf{D}(\tau)=\text{diag}\,(0,0,0)$ or $\text{diag}\,\varsigma\pr{-\lambda_1^2(\tau),-\lambda_1^2(\tau),0}$.
The last two results can be summarized by saying that when $\nsd=2$,
\begin{equation}
\usf{D}(\tau)=\text{diag}\pr{-\lambda^2(\tau),-\lambda^2(\tau)},
\label{eq:DExpFormnsd2}
\end{equation}
where $\lambda(\tau)\in\m{R}$, and when $\nsd=3$, $\usf{D}(\tau)=\text{diag}\,\varsigma(-\lambda(\tau)^2,-\lambda(\tau)^2,0)$.
Without loss of generality, we can choose the order of $\usf{n}_i(\tau)$ so that their respective eigenvalues form a non-increasing sequence\footnote{The matrix $\busf{W}(\tau)$ as the square root of $\text{sym}\pr{\usf{P}(\tau)}$ does not depend on the order of $\usf{n}_i(\tau)$. Different orders
will lead to the same $\busf{W}(\tau)$.}.
Therefore, for concreteness in the case of $\nsd=3$ we take
\begin{align}
\usf{D}(\tau)&=\begin{pmatrix}0 & 0 & 0\\0 & -\lambda^2(\tau) & 0\\0 &0 & -\lambda^2(\tau)\end{pmatrix}.
\label{eq:DExpFormnsd3}
\end{align}

\subsection{Calculation of $\busf{W}(\tau)$ from $\text{sym}\pr{\usf{P}(\tau)}$ using $\eqref{equ:sym-1}$\label{subsec:Calculation-of-}}

Let
\begin{equation}
\usf{F}(\tau):=\tpsb{\usf{N}}(\tau)\,\busf{W}(\tau)\,\usf{N}(\tau).\label{eq:FDef}
\end{equation}
It can be shown using $\usf{F}(\tau)$'s definition, equations $\eqref{eq:NOrthonormalityCondition}$, and $\eqref{equ:sym-1}$,
and $\text{sym}\pr{\usf{P}(\tau)}$'s decomposition that is derived in \ref{subsec:A-spectral-decomposition} and summarized in $\eqref{equ:SymPTheoreticalSpecDecomposition}$ that

\begin{equation}
\usf{F}^2(\tau)=\usf{D}(\tau).
\label{eq:F2D}
\end{equation}
Substituting $\usf{D}(\tau)$ in $\eqref{eq:F2D}$ from $\eqref{eq:DExpFormnsd2}$ and $\eqref{eq:DExpFormnsd3}$ for then noting from Lemma \ref{lemma:skew} and $\usf{F}(\tau)$'s definition that $\usf{F}(\tau)$ is skew-symmetric, it can be shown that for
$\nsd=2$ and $3$
\begin{equation}
\usf{F}(\tau)=
\pm
\star\pr{\lambda(\tau)}
\label{equ:FfromLambda2}
\end{equation}
and
\begin{equation}
\usf{F}(\tau)=
\pm
\star\pr{\pr{\lambda(\tau),0,0}},
\label{equ:FfromLambda3}
\end{equation}
respectively.
Equations $\ref{equ:WbTheoreticalDecomposition}$ follow from $\eqref{eq:FDef}$, $\eqref{equ:FfromLambda2}$, and $\eqref{equ:FfromLambda3}$.

\section*{References}
\bibliography{mybibfile}

\end{document}